\documentclass[fleqn,usenatbib]{mnras}

\usepackage{newtxtext,newtxmath}

\usepackage[T1]{fontenc}
\usepackage{ae,aecompl}

\usepackage{graphicx}	
\usepackage{amsmath}	
\usepackage{amssymb}	

\title[Dust in high-$z$ galaxies]
{Radiative equilibrium estimates of dust temperature and mass in high-redshift galaxies}

\author[A.\ K.\ Inoue et al.]{
  Akio K.\ Inoue,$^{1,2,3}$\thanks{E-mail: akinoue@aoni.waseda.jp (AKI)}
  Takuya Hashimoto,$^{4,5,6,3}$
  Hiroki Chihara,$^{3}$
  and Chiyoe Koike$^{3}$
\\
$^{1}$Department of Physics, School of Advanced Science and Engineering, Faculty of Science and Engineering, Waseda University, 3-4-1, Okubo, Shinjuku, Tokyo\\ 169-8555, Japan\\
$^{2}$Waseda Research Institute for Science and Engineering, Faculty of Science and Engineering, Waseda University, 3-4-1, Okubo, Shinjuku, Tokyo 169-8555, Japan\\
$^{3}$Department of Environmental Science and Technology, Faculty of Design Technology, Osaka Sangyo University, 3-1-1, Nakagaito, Daito, Osaka 574-8530, Japan\\
$^{4}$Tomonaga Center for the History of the Universe (TCHoU),
Faculty of Pure and Applied Sciences, University of Tsukuba,
Tsukuba, Ibaraki 305-8571, Japan\\
$^{5}$Faculty of Science and Engineering, Waseda University, 3-4-1, Okubo, Shinjuku, Tokyo 169-8555, Japan\\
$^{6}$National Astronomical Observatory of Japan, 2-21-1, Osawa, Mitaka, Tokyo 181-8588, Japan
}

\date{Accepted XXX. Received YYY; in original form ZZZ}

\pubyear{2019}

\begin{document}
\label{firstpage}
\pagerange{\pageref{firstpage}--\pageref{lastpage}}
\maketitle

\begin{abstract}
Estimating the temperature and mass of dust in high-$z$ galaxies is essential for discussions of the origin of dust in the early Universe.
However, this suffers from limited sampling of the infrared spectral-energy distribution.
Here we present an algorithm for deriving the temperature and mass of dust in a galaxy, assuming dust to be in radiative equilibrium.
We formulate the algorithm for three geometries: a thin spherical shell, a homogeneous sphere, and a clumpy sphere.
We also discuss effects of the mass absorption coefficients of dust at ultraviolet and infrared wavelengths, $\kappa_{\rm UV}$ and $\kappa_{\rm IR}$, respectively.
As an example, we apply the algorithm to a normal, dusty star-forming galaxy at $z=7.5$, A1689zD1, for which three data points in the dust continuum are available.
Using $\kappa_{\rm UV}=5.0\times10^4$ cm$^2$ g$^{-1}$ and $\kappa_{\rm IR}=30(\lambda/100\micron)^{-\beta}$ cm$^2$ g$^{-1}$ with $\beta=2.0$, we obtain dust temperatures of 38--70~K and masses of $10^{6.5-7.3}$ M$_\odot$ for the three geometries considered.
We obtain similar temperatures and masses from just a single data point in the dust continuum, suggesting the usefulness of the algorithm for high-$z$ galaxies with limited infrared observations.
In the clumpy-sphere case, the temperature becomes equal to that of the usual modified black-body fit, because an additional parameter describing the clumpiness works as an adjuster.
The best-fit clumpiness parameter is $\xi_{\rm cl}=0.1$, corresponding to $\sim10$\% of the volume filling factor of the clumps in this high-$z$ galaxy if the clump size is $\sim10$ pc, similar to that of giant molecular clouds in the local Universe.
\end{abstract}




\section{Introduction}

Solid particles in interstellar space--called \lq\lq cosmic dust\rq\rq--are ubiquitous.
Even in the era of the first objects in the Universe, dust grains may exist if the first objects were massive stars that ended as supernovae that produced such grains \citep{Nozawa03}.
Observationally, infrared (IR) thermal emission from dust has already been detected from
star-forming galaxies in the early Universe at redshifts $z>7$ \citep{Watson15,Laporte17,Tamura18,Hashimoto19}.\footnote{We restrict our discussion to the dust in star-forming galaxies, not in QSOs.}
The estimated mass of dust in these galaxies is as large as $\sim10^{6-7}$ M$_\odot$, and the dust-to-stellar mass ratio reaches $\sim10^{-2}$, which is an order of magnitude larger than the median value for local galaxies \citep{Calura17}. 
Such large amounts of dust require efficient growth of the dust mass in dense clouds in the interstellar medium (ISM), because of insufficient dust production by supernovae, which is the unique path for stellar dust production at the early times before asymptotic giant branch stars appear (e.g. \citealt{Michalowski15}).
On the other hand, many high-$z$ galaxies have not yet been detected in the IR dust continuum, indicating that they contain significantly less dusty than their dusty counterparts.
For example, \cite{Hashimoto18} reported a dust-to-stellar mass ratio of $<10^{-4}$ in a $z=9.11$ galaxy, more than an order of magnitude smaller than the local median value \citep{Calura17}.
Thus, there seems to be a large diversity in the amounts of dust in high-$z$ galaxies.

However, the estimated dust mass in high-$z$ galaxies may suffer from large uncertainties 
because of the unknown dust temperature--which is required to obtain the mass--in addition to large uncertainties in the IR emissivity of the dust.
Even for the detected galaxies, only one or two data points are available in the IR, except for sufficiently bright sub-millimetre galaxies for which the IR spectral energy distribution (SED) is well sampled \citep{Riechers13,Marrone18}.
It is difficult to determine the temperature from the sparse data points available for the SEDs of normal dusty galaxies (e.g. \citealt{Capak15,Watson15}).
Therefore, a temperature of 40--50 K has often been assumed in the literature \citep{Tamura18,Hashimoto18,Hashimoto19}.
This choice is based on observations of such a high dust temperature found in low-$z$ galaxies which have properties similar to the high-$z$ ones \citep{Faisst17}.
This is in contrast to local galaxies like the Milky Way, for which the dust temperature is typically 16--18 K (e.g. \citealt{Okumura96}).

A possible solution to this problem is to use IR SED templates.
A number of empirical or theoretical templates for the SEDs due to IR dust emission have been proposed to date (e.g. \citealt{CharyElbaz01,TotaniTakeuchi02,DaleHelou02,DraineLi07,Rieke09,Casey12,Dale14}).
The simplest models have only a single parameter, such as the total IR luminosity or the dust temperature.
The IR luminosity can be equated to the ultraviolet (UV)-to-optical luminosity absorbed by the dust, which in turn can be estimated from UV-to-optical SED fits, or more simply from UV spectral slopes.
However, the latter approach suffers from the large uncertainty of the so-called \lq\lq IRX-$\beta$ relation\rq\rq\footnote{IRX stands for InfraRed eXcess and is defined as the luminosity ratio of IR to UV. $\beta$ is the spectral index of the flux density and is defined as $F_\lambda\propto\lambda^{\beta}$.} (e.g. \citealt{Meurer99,Buat05,Takeuchi12,Faisst17}).
Even using SED fitting codes like {\sc magphys} \citep{daCunha08} or {\sc cigale} \citep{Boquien19}, in which the dust absorption and emission are treated in a self-consistent way, the limited sampling of the IR SEDs presents difficulties for high-$z$ galaxies.
\cite{Hirashita17} examined the entire SED including an upper limit at a sub-mm wavelength of {\it Himiko} \citep{Ouchi09}, which yielded only an upper limit to the dust mass as a function of dust temperature.

Any methods based on IR SED templates necessarily rely on the applicability of those templates to the sample galaxies.
However, for high-$z$ galaxies, the IR SEDs have not yet been explored in detail.
\cite{Casey18} reported a good empirical correlation from $z\sim0$ to $z\sim5$ between the peak wavelength of the IR SED (or, equivalently, the dust temperature) and the total IR luminosity.
This may support the applicability of IR SED templates constructed in the local Universe to high-$z$ galaxies.
On the other hand, a different situation has been found in the latest numerical simulations that predict the IR SEDs of high-$z$ galaxies \citep{Narayanan18,Behrens18,Arata19,Ma19,Liang19}.
For galaxies at $z>6$, \cite{Ma19} predict a systematically shorter peak wavelength for a given luminosity than that found by \cite{Casey18}.
\cite{Arata19} found the same result in terms of the dust temperature--IR luminosity relation.
These results caution us against applying empirical low-$z$ IR SED templates uncritically to very high-$z$ galaxies.

Here we propose another way to estimate the dust temperature and mass: a radiative-equilibrium method.
The IR luminosity of the dust originates from the luminosity it absorbs.
If we know the absorbed luminosity, radiative equilibrium determines the thermal-emission temperature of the dust.
\cite{Hirashita14} employed such an approach to estimate the temperature and the corresponding mass of dust by using the observed UV luminosity as well as the IR upper limit for {\it Himiko} \citep{Ouchi09}.
However, \cite{Hirashita14} adopted an optically thin and geometrically thin dust-shell geometry in their formulation, and they did not treat any radiative-transfer effects.
In this paper, we expand their approach by taking radiative transfer into account, and we formulate the algorithm for three geometries: a geometrically thin spherical shell, a homogeneous sphere, and a clumpy sphere.
Thanks to these simple geometries, we can solve the radiative-transfer equations analytically (see also \citealt{Imara18}).
We expect these analytic formulae to be useful for future applications to a large set of IR observations of galaxies, and it is also easy to implement them in SED fitting codes.
As an example, we apply our algorithm to the high-$z$ galaxy A1689zD1 and estimate the temperature and mass of the dust in this galaxy.
We also discuss briefly the effect of the IR emissivity of the dust on estimates of the dust temperature and mass.

The rest of this paper is structured as follows: 
In section~2, we present our own observational data from the Atacama Large Millimetre/submillimetre Array (ALMA) for the galaxy A1689zD1.
The full formulae for our algorithm are given in section~3. 
In section~4, we present as an example the application of the algorithm to A1680zD1, and we compare the results with the common method of estimating the dust temperature by using a modified black-body fit.
The final section is devoted to a summary of our findings.
We assume the cosmological parameters to be $H_0=70$ km s$^{-1}$ Mpc$^{-1}$, $\Omega_{\rm m}=0.3$, and $\Omega_\Lambda=0.7$.
The definition of the AB magnitude is found in \cite{OkeGunn83}.

\section{The sample galaxy}

In this paper, we apply our algorithm, described in the next section, to a dusty star-forming galaxy at $z\simeq7.5$, A1689zD1 \citep{Watson15}, in order to demonstrate the validity and usefulness of the algorithm.
In this section, we present detailed information about A1689zD1 and our ALMA Band~8 observations.

\subsection{A normal, dusty star-forming galaxy, A1689zD1}

A1689zD1 was discovered by \citet{Bradley08} as an apparently bright ($m\sim25$ AB) $z>7$ galaxy candidate thanks to strong gravitational-lensing (the magnification factor is $\mu_{\rm GL}\simeq9$; \citealt{Bradley08}) by the foreground galaxy cluster, A1689, at $z=0.1832$ \citep{Struble99}.
The de-lensed magnitude of this galaxy is $m\sim27$, which corresponds to a star-formation rate (SFR) of $\sim 6~M_\odot$~yr$^{-1}$, using Kennicutt's conversion \citep{Kennicutt98}.
Thus, this galaxy is much less active than sub-millimetre galaxies and even Lyman-break galaxies (see \citealt{Watson15}).
The galaxy was detected at the wavelength of 1.3~mm in ALMA Band~6, being the first discovery of dust beyond $z=7$ \citep{Watson15}.
Follow-up ALMA observations in Band~7 brought another significant continuum detection at 0.87~mm \citep{Knudsen17}. 
We have also made follow-up ALMA observations in Band~8 and have detected the dust continuum at 0.73~mm, as described in the next subsection.
There are thus three continuum detections of A1689zD1, making it an ideal galaxy to test our method for estimating the dust temperature and mass in $z>7$ galaxies.

As preparation for the following sections, we here estimate the escaping UV luminosity that is transmitted through the ISM of the galaxy, defined as $L_{\rm UV}^{\rm esc}=\nu_{\rm UV}L_{\nu_{\rm UV}}^{\rm esc}=4\pi d_{\rm L}^2 (c/\lambda_{\rm obs}) F_{\nu_{\rm obs}}$, where $\nu_{\rm UV}$ is the UV frequency corresponding to the observing wavelength $\lambda_{\rm obs}$, $F_{\nu_{\rm obs}}$ is the observed (de-lensed) flux density (per unit frequency) at $\lambda_{\rm obs}$, $d_{\rm L}$ is the luminosity distance, and $c$ is the speed of light.
The flux density is obtained from the observed magnitude $m_{H140}=24.64\pm0.05$ AB \citep{Watson15}, for which the corresponding wavelength is $\approx1600$ \AA\ in the rest-frame of the galaxy.
Taking into account a conservative redshift uncertainty of $\pm0.4$ (a $2\sigma$ range, from the report by \citealt{Watson15}) as well as the photometric uncertainty, we obtain 
$L_{\rm UV}^{\rm esc}=(8.0\pm1.0)\times(9/\mu_{\rm GL})\times10^{43}$ erg s$^{-1}$.

In Table~\ref{tab:obsdata}, we summarise the observational data for A1689zD1.

\begin{table*}
  \centering
  \caption{A summary of the observational properties of A1689zD1.}
  \label{tab:obsdata}
  \begin{tabular}{llll}
    \hline
    \multicolumn{2}{l}{Basic properties} & Remarks & Reference \\
    \hline
    RA & 13$^h$11$^m$29.96$^s$ & J2000 & NASA NED \\
    Dec & $-$01$^d$19$^m$18.7$^s$ & J2000 & NASA NED \\
    $z$ & $7.5\pm0.2$ & Ly$\alpha$ break & Watson et al.~(2015) \\
    $m_{H140}$ & $24.64\pm0.05$ & AB & Watson et al.~(2015) \\
    $\mu_{\rm GL}$ & 9 & Fiducial value & Watson et al.~(2015) \\
    $L_{\rm UV}^{\rm esc}$ & $(8.0\pm1.0)\times10^{43}$ & $(9/\mu_{\rm GL})$ erg s$^{-1}$ & This work \\
    $SFR_{\rm UV}$ & 6.3 & M$_\odot$ yr$^{-1}$ & This work \\
    \hline
    \multicolumn{4}{l}{Dust continuum (observed)} \\
    \hline
    & $\lambda$ [mm] & $F_\nu^{\rm obs}$ [mJy] & Reference \\
    \hline
    Band 8 & $0.728$ & $1.67\pm0.36$ & This work \\
    Band 7 & $0.873$ & $1.33\pm0.14$ & Knudsen et al.~(2017) \\
    Band 6 & $1.33$ & $0.56\pm0.10$ & Watson et al.~(2015) \\
    \hline
    \multicolumn{4}{l}{Size information} \\
    \hline
    & Band~8 & Band~7 & Remarks \\
    \hline
    $a$ & $1.''45\pm0.''35$ & $1.''67\pm0.''23$ & FWHM along the major axis \\
    $b$ & $0.''60\pm0.''18$ & $0.''51\pm0.''11$ & FWHM along the minor axis \\
    $R_{\rm o}$ & $0.''47\pm0.''09$ & $0.''46\pm0.''06$ & $\sqrt{ab}/2$ (observed) \\
    $R$ & $0.78 \pm 0.15$ & $0.77 \pm 0.10$ & $(3/\sqrt{\mu_{\rm GL}})$ kpc (proper) \\
    \hline
  \end{tabular}
\end{table*}

\subsection{ALMA Band 8 observations}

We obtained the ALMA Band 8 observations for A1689zD1 (2016.1.00954.S; PI: A.~K.~Inoue) in November 2016, using 40 antennas with baseline lengths of 15--704 m in the dual-polarization setup. The total on-source exposure time was 48 minutes. 
We used four spectral windows, with bandwidths of 1.875 GHz, in the Frequency Division Mode with a channel spacing of 7.8125 MHz. The lower and upper sidebands covered the contiguous frequency ranges of 393.49--397.16 GHz and 405.55--409.22 GHz, respectively. 
We used quasar J1256-0547 for bandpass and flux calibrations and employed quasar J1312-0424 for phase calibrations. We estimate the flux-calibration uncertainty to be better than 10\%. We reduced and calibrated the data using the Common Astronomy Software Applications (CASA; \citealt{McMullin07}), pipeline version 4.7.0. We produced images with the {\tt CLEAN} task using natural weighting with a 0.5 arcsec taper.
With the CASA task {\tt imstat}, we estimated the rms level of the continuum image to be 77 $\mu$Jy beam$^{-1}$. The spatial resolution of the continuum image is $0''.61 \times 0''.53$ in FWHM with the beam position angle PA $= 65^{\circ}$. 

\begin{figure}
  \centering
  \includegraphics[width=7cm]{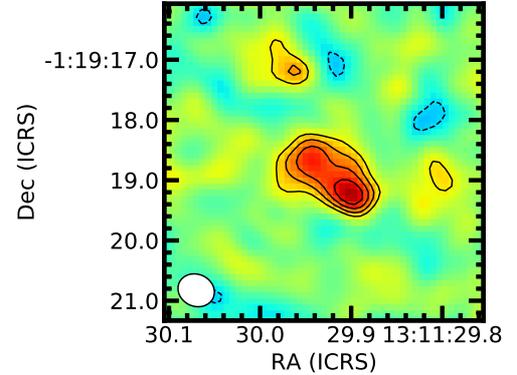}
  \caption{ALMA Band~8 continuum image of A1689zD1. A $0.''5$ taper has been applied. The contours represent $-2, 2, 3, 4$, and $5\sigma$ flux densities with $\sigma=77$ $\mu$Jy beam$^{-1}$. Negative contours are depicted by dashed lines. The beam size and position angle, as indicated by the white ellipse in the bottom left corner, are $0.''61 \times 0.''53$ and $65^\circ$, respectively.}
  \label{fig:Band8}
\end{figure}

Figure~\ref{fig:Band8} shows the dust continuum emission detected at the position of A1689zD1. 
By spatially integrating the image using the CASA task {\tt imfit} with a 2D Gaussian profile, we estimate the continuum flux density to be $1.67 \pm 0.36$ mJy. 
The apparent (or lensed) beam-deconvolved size is $(1.45\pm0.35) \times (0.60\pm0.18)$ arcsec$^{2}$ at PA $=62 \pm 11^{\circ}$.
Comparison with the Band~6 and 7 observations by \citet{Watson15} and \citet{Knudsen17}, respectively, shows that the spatial position and extension of the Band~8 continuum are quite consistent with those in Bands~6 and 7.
Quantitatively, we have measured the apparent beam-deconvolved size in the archived Band~7 data to be $(1.67\pm0.23) \times (0.51\pm0.11)$ arcsec$^{2}$ at PA $=54 \pm 4^{\circ}$.
Thanks to the higher S/N ratio in Band~7, the uncertainties are smaller than those in Band~8.
There may be two components to the north-east and south-west in Band~6, as reported by \citet{Knudsen17}.
Nevertheless, in the following sections, we assume for simplicity that the galaxy is a single sphere.
For the radius $R$ of the sphere, we used the geometric mean of the semi-major and semi-minor axes of the Band~7 continuum: $R_{\rm o}=0.''46\pm0.''06$. This corresponds to a physical scale of $R=0.77\pm0.10\times(3/\sqrt{\mu_{\rm GL}})$ kpc, if we assume symmetric lensing magnification.
This size is quite consistent with the mean effective radius in the rest-frame UV for $z\sim6$--7 galaxies reported by \citet{Kawamata18}, given the de-lensed absolute magnitude of the object ($M_{\rm UV}=-20.2$).

\section{Algorithm for determining the dust mass and temperature}

In this section, we describe our algorithm for determining the 
temperature and mass of the dust under the assumption of radiative equilibrium. 

\subsection{Definitions}

First, we define the mean mass absorption coefficient of the dust grains as 
\begin{equation}
    \kappa_\nu \equiv \frac{\int \sigma_\nu(a) n'_{\rm d}(a) da}{\int m_{\rm d}(a) n'_{\rm d}(a) da}
    = \frac{3 \langle a^2 Q_\nu \rangle}{4s \langle a^3 \rangle}\,.
    \label{MACdef}
\end{equation}
The symbols $\sigma_\nu(a)$ and $m_{\rm d}(a)$ are the absorption cross-section at frequency $\nu$ and the mass of a grain with radius $a$, respectively.
The term $n'_{\rm d}(a) da$ is the number density of grains with radii in the range $a$ to $a+da$; i.e. the size-distribution function.
The total number density of grains is given by $n_{\rm d}=\int n'_{\rm d}(a)da$.
For compact spherical grains, the cross-section becomes $\sigma_\nu(a)=\pi a^2 Q_\nu(a)$, where $Q_\nu$ is the absorption Q-parameter, and the mass becomes $m_{\rm d}(a)=(4\pi/3) s a^3$, with $s$ being the density of the dust material. For the last expression, we have introduced the following two averages over the grain-size distribution: 
$\langle a^2 Q_\nu \rangle = \int a^2 Q_\nu(a) n'_{\rm d}(a) da / n_{\rm d}$ and 
$\langle a^3 \rangle = \int a^3 n'_{\rm d}(a) da / n_{\rm d}$.
Figure~\ref{fig:kappa} shows examples of $\kappa$ as a function of wavelength for some dust models. 

Next, we define the dust temperature, which we discuss in this paper. 
If we neglect self-absorption of the dust emission, which is reasonable at IR wavelengths on the scale of a galaxy (but see also \citealt{Ferrara17}), the thermal luminosity of the dust can be expressed as 
\begin{equation}
    L_{\rm d}^{\rm em} = \int \int \int 4 \pi B_\nu(T_{\rm d}[a]) 
    \sigma_\nu(a) n'_{\rm d}(a) da d\nu dV\,,
    \label{LIR}
\end{equation}
where $B_\nu$ is the Planck function, with $T_{\rm d}(a)$ being the radiation temperature of grains of radius $a$, and $\int dV$ means integration over the volume of the galaxy. 
In general, the dust temperature depends upon the location or environment within the galaxy as well as upon the grain size and material. Dealing with these effects requires solving radiation-transfer equations in the galaxy, which is a complex task with a number of degrees of freedom. However, in this paper,
we assume a single dust temperature to represent the distribution of temperatures of the grains in the galaxy. 
Fortunately, since we will deal with rest-frame far-infrared (FIR) observations, 
the SED of the galaxy can be approximated very well by a modified Planck function with a single temperature. 
The temperature can be regarded as a luminosity-weighted average temperature of grains and be biased towards the temperature of rather small and warm grains.
In addition, to make the problem tractable analytically we assume a uniform ISM. In this case, equation~(\ref{LIR}) can be simplified to
\begin{equation}
    L_{\rm d}^{\rm em} = 4\pi M_{\rm d} \int \kappa_\nu B_\nu(T_{\rm d}) d\nu\,,
    \label{Ldust}
\end{equation}
where $M_{\rm d}=\int (4\pi/3) s \langle a^3 \rangle n_{\rm d} dV$ is the total mass of dust in the galaxy, $\kappa_\nu$ is the mass absorption coefficient defined in equation~(\ref{MACdef}), and $T_{\rm d}$ is the representative dust temperature in the galaxy.

Another point to be clarified is that we will discuss the ``intrinsic'' dust radiation temperature, including the cosmic microwave background (CMB) heating, which depends upon the redshift. 
This is different from the temperature corrected for CMB heating, which would be observed if the galaxy were located at $z=0$ \citep{daCunha13}.

\subsection{Radiative equilibrium and temperature}

The radiative equilibrium of the dust grains can be expressed as 
\begin{equation}
 L^{\rm em}_{\rm d}(T_{\rm d}) = L^{\rm abs}_* + L^{\rm abs}_{\rm CMB}\,,
 \label{Radeq}
\end{equation}
where $L^{\rm abs}_*$ is the stellar luminosity absorbed by the dust grains, and $L^{\rm abs}_{\rm CMB}$ is the CMB luminosity absorbed by the dust grains. 
The luminosity emitted by the dust, $L_{\rm d}^{\rm em}$, is given by equation~(\ref{Ldust}).
For the specific emissivity $\kappa_\nu$ in the IR (i.e. the mass absorption coefficient), 
we employ the approximation $\kappa_\nu=\kappa_{\rm IR,0}(\nu/\nu_0)^\beta$, 
where $\beta$ is the emissivity index, and $\kappa_{\rm IR,0}$ is the pivot emissivity at the frequency $\nu_0$. 
Then we obtain analytically 
\begin{equation}
 L^{\rm em}_{\rm d} = C \kappa_{\rm IR,0} M_{\rm d} T_{\rm d}^{\beta+4}\,,
 \label{analyticLd}
\end{equation}
and 
\begin{equation}
 C = \frac{8\pi k_{\rm B}^{\beta+4}}{c^2 \nu_0^\beta h^{\beta+3}}
  \zeta(\beta+4) \Gamma(\beta+4)\,,
\end{equation}
where $h$ is Planck's constant, $c$ is the speed of light in the vacuum,
$k_{\rm B}$ is Boltzmann's constant, $\zeta$ is the Zeta function, and
$\Gamma$ is the Gamma function. The absorbed CMB luminosity is
also given by equation~(\ref{analyticLd}) if $T_{\rm d}$ is replaced by the CMB temperature $T_{\rm CMB}$: 
$L^{\rm abs}_{\rm CMB}=C \kappa_{\rm IR,0} M_{\rm d} T_{\rm CMB}^{\beta+4}$.
Again, we have assumed the medium to be optically thin for the CMB
radiation. Substituting equation~(\ref{analyticLd}) into equation~(\ref{Radeq}), we obtain 
\begin{equation}
 T_{\rm d} = \left(\frac{L^{\rm abs}_*}{C \kappa_{\rm IR,0} M_{\rm d}}
	     + T_{\rm CMB}^{\beta+4}\right)^{\frac{1}{\beta+4}}\,.
 \label{TdEq}
\end{equation}

\subsection{Effective optical depth and escape probability}

We assume that the stellar radiation energy is dominated by UV
radiation, which is reasonable in star-forming galaxies observed at
high-$z$. We also assume that the intrinsic UV luminosity is equal to
the sum of the escaping (i.e. observed) UV luminosity 
$L_{\rm UV}^{\rm esc}$ and the stellar luminosity absorbed by the dust, $L_*^{\rm abs}$. 
Using the escape probability of UV radiation from a medium, 
$P_{\rm esc}(\tau)$, where $\tau$ is the effective optical depth of the medium, 
we can express the absorbed luminosity as 
\begin{equation}
 L_*^{\rm abs} = L_{\rm UV}^{\rm esc} \frac{1-P_{\rm esc}(\tau)}{P_{\rm esc}(\tau)}\,.
 \label{eq:Labs}
\end{equation}
In the following, we present analytic expressions for the effective
optical depth $\tau$ and the escape probability $P_{\rm esc}(\tau)$
for three simple geometries.
\cite{Imara18} have presented similar solutions for shell and homogeneous geometries.
For simplicity, we neglect scattering in this paper.

\subsubsection{Spherical shell}

Consider a uniform {\it thin} spherical shell of dust grains that
surrounds radiation sources. The mass of dust is $M_{\rm d}$, and the
radius of the shell is $R$. The column density of the dust mass in the shell is
$\Sigma_{\rm d}=M_{\rm d}/4\pi R^2$, and the optical depth for UV radiation is $\tau_{\rm she}=\kappa_{\rm UV} \Sigma_{\rm d}$, 
where $\kappa_{\rm UV}$ is the mean mass absorption coefficient for the UV (see equation~\ref{MACdef} and section~\ref{subsec:dust}).
We therefore obtain 
\begin{equation}
 \tau_{\rm she}=\frac{\kappa_{\rm UV} M_{\rm d}}{4\pi R^2}\,.
 \label{eq:taushe}
\end{equation}
The escape probability is simply 
\begin{equation}
 P_{\rm esc}^{\rm she}(\tau)=e^{-\tau}\,.
 \label{eq:P_shell}
\end{equation}

\subsubsection{Homogeneous sphere}

Next, consider a spherical medium in which dust grains and radiation sources are
distributed uniformly. Again, the dust mass is $M_{\rm d}$ and the radius
of the sphere is $R$. 
The UV optical depth of the medium in the radial direction is
$\tau_{\rm hom}=\kappa_{\rm UV} \rho_{\rm d} R$, 
where $\rho_{\rm d}$ is the mass density of the dust,
which is given by $\rho_{\rm d}=3M_{\rm d}/4\pi R^3$.
Therefore we obtain 
\begin{equation}
 \tau_{\rm hom}=\frac{3\kappa_{\rm UV}M_{\rm d}}{4\pi R^2}\,.
 \label{eq:tauhom}
\end{equation}
Note that $\tau_{\rm hom}$ is 3 times larger than $\tau_{\rm she}$,
because some grains can be closer to the radiation sources, so the 
solid angle they subtend, as seen by the sources, becomes larger than in the shell case. The
escape probability from a homogeneous sphere of optical depth 
$\tau$ is given by (e.g. \citealt{Osterbrock89} and Appendix C of
\citealt{Varosi99}) 
\begin{equation}
 P_{\rm esc}^{\rm hom}(\tau) = \frac{3}{4\tau} 
  \left\{
   1-\frac{1}{2\tau^2}
   +\left(\frac{1}{\tau}+\frac{1}{2\tau^2}\right)
   e^{-2\tau}
  \right\} \,.
  \label{eq:P_hom}
\end{equation}

\subsubsection{Clumpy sphere: Mega-grain approximation}\label{sec:clumpy}

The ``Mega-grain'' approximation is an analytical treatment of radiation transfer in a clumpy medium, 
where the sizes of the clumps are small compared to the system size (e.g. \citealt{Neufeld91,Hobson93,Varosi99,Inoue05}). 
In this approximation, the clumps can be regarded as huge dust grains, called ``Mega-grains,'' which absorb (and scatter) radiation. 
In the equations, we can just replace the usual single-grain opacity with an effective opacity of the clumps. 

Consider a spherical medium of radius $R$ that consists of clumps and the inter-clump medium.
We assume the clumps all to be identical, with radius $r_{\rm cl}$, and to be distributed uniformly throughout the sphere.
The radiation sources are also assumed to be distributed uniformly and not to be correlated with the distribution of the clumps. 
Again, we take the total mass of dust in the system to be $M_{\rm d}$ and the mean mass density of the dust to be $\rho_{\rm d}=3M_{\rm d}/4\pi R^3$.
We denote the dust densities in the clumps and in the inter-clump medium by $\rho_{\rm d,cl}$ and $\rho_{\rm d,ic}$, respectively.
If the volume fraction of clumps is $f_{\rm cl}$, 
the mean density is $\rho_{\rm d}=f_{\rm cl}\rho_{\rm d,cl}+(1-f_{\rm cl})\rho_{\rm d,ic}$.
If we define the density contrast between the clumps and the inter-clump medium to be $C_{\rm cl}=\rho_{\rm d,cl}/\rho_{\rm d,ic}\geq1$, then the two densities become 
\begin{equation}
    \rho_{\rm d,cl}=\frac{C_{\rm cl}\rho_{\rm d}}{(C_{\rm cl}-1)f_{\rm cl}+1}\,,
\end{equation}
and 
\begin{equation}
    \rho_{\rm d,ic}=\frac{\rho_{\rm d}}{(C_{\rm cl}-1)f_{\rm cl}+1}\,.
\end{equation}
In the limiting cases, (1) $\rho_{\rm d,cl}\to \rho_{\rm d}/f_{\rm cl}$ and $\rho_{\rm d,ic}\to0$ when $C_{\rm cl}\to\infty$, and 
(2) $\rho_{\rm d,cl}=\rho_{\rm d,ic}=\rho_{\rm d}$ when $C_{\rm cl}=1$ (i.e. the homogeneous sphere case).

In the Mega-grain approximation, the effective optical depth of the system can be expressed as 
\begin{equation}
    \tau_{\rm eff}=\tau_{\rm MG} + \tau_{\rm ic}\,,
    \label{eq:taueff}
\end{equation}
where $\tau_{\rm MG}$ and $\tau_{\rm ic}$ are the Mega-grain optical depth and the inter-clump optical depth, respectively.
The latter is simply given by $\tau_{\rm ic}=\kappa_{\rm UV} \rho_{\rm d,ic} R$, where the UV mass absorption coefficient is $\kappa_{\rm UV}$.
The former is given by $\tau_{\rm MG}=n_{\rm cl} \sigma_{\rm cl} R$, where the number density of clumps is $n_{\rm cl}=(3f_{\rm cl})/(4\pi r_{\rm cl}^3)$, and the radiation cross-section of a single clump is $\sigma_{\rm cl}=\pi r_{\rm cl}^2 Q_{\rm cl}$, where $Q_{\rm cl}$ is the (absorption) ``$Q$-parameter'' for a single clump.
For a spherical clump of radial optical depth $\tau_{\rm cl}$, we find 
\begin{equation}
    Q_{\rm cl}=\frac{4\tau_{\rm cl}}{3}P_{\rm esc}^{\rm hom}(\tau_{\rm cl})\,
\end{equation}
(Appendix C of \citealt{Varosi99}).
The radial optical depth of a single clump is 
\begin{equation}
    \tau_{\rm cl} = \kappa_{\rm UV} (\rho_{\rm d,cl} - \rho_{\rm d,ic}) r_{\rm cl} 
    = \tau_{\rm hom} 
    \left\{\frac{(C_{\rm cl}-1)\eta_{\rm cl}}
    {(C_{\rm cl}-1)f_{\rm cl}+1}\right\}\,,
\end{equation}
where $\eta_{\rm cl}=r_{\rm cl}/R$. 
For the limiting cases, (1) $\tau_{\rm cl}\to\tau_{\rm hom}(\eta_{\rm cl}/f_{\rm cl})$ when $C_{\rm cl}\to\infty$, and (2) $\tau_{\rm cl}=\tau_{\rm hom}$ when $C_{\rm cl}=1$.
Note that we must subtract the inter-clump optical depth, because it is already taken into account in equation~(\ref{eq:taueff}).

Finally, the effective optical depth (equation~\ref{eq:taueff}) becomes
\begin{equation}
    \tau_{\rm eff}=\tau_{\rm hom}
        \frac{(C_{\rm cl}-1)f_{\rm cl}P_{\rm esc}^{\rm hom}(\tau_{\rm cl})+1}
        {(C_{\rm cl}-1)f_{\rm cl}+1}\,.
\end{equation}
In the limiting cases, (1) $\tau_{\rm eff}\to\tau_{\rm hom}P_{\rm esc}^{\rm hom}(\tau_{\rm cl})$ when $C_{\rm cl}\to\infty$, and (2) $\tau_{\rm eff}=\tau_{\rm hom}$ when $C_{\rm cl}=1$ (i.e. for a homogeneous medium).
Since the clump distribution is uniform, the escape probability from the clumpy sphere is the same as for the homogeneous case, but with the effective optical depth $\tau_{\rm eff}$: 
\begin{equation}
 P_{\rm esc}^{\rm MG}=P_{\rm esc}^{\rm hom}(\tau_{\rm eff})\,.
\end{equation}
In the following, we consider only the high-contrast limit (i.e. $C_{\rm cl}\gg1$).
Namely, 
\begin{equation}
    \tau_{\rm cl} \approx \tau_{\rm hom} \xi_{\rm cl}\,,
\end{equation}
where we have introduced the \lq\lq clumpiness parameter\rq\rq\ $\xi_{\rm cl}=\eta_{\rm cl}/f_{\rm cl}$, and 
\begin{equation}
    \tau_{\rm eff} \approx \tau_{\rm hom} P_{\rm esc}^{\rm hom}(\tau_{\rm cl})\,.
\end{equation}

For the discussions in section~\ref{Overall} below, we note here some limiting cases for clumpy media described by the clumpiness parameter $\xi_{\rm cl}=\eta_{\rm cl}/f_{\rm cl}$.
Recalling that $\eta_{\rm cl}=r_{\rm cl}/R$, and introducing the total number of clumps $N_{\rm cl}$, we find the volume filling factor to be $f_{\rm cl}=N_{\rm cl}\eta_{\rm cl}^3$, and $\xi_{\rm cl}=1/(N_{\rm cl}\eta_{\rm cl}^2)$.
In addition, $\eta_{\rm cl}\leq1/N_{\rm cl}^{1/3}$, since $f_{\rm cl}\leq1$ by definition, and then $\xi_{\rm cl}\geq1/N_{\rm cl}^{1/3}\geq\eta_{\rm cl}$.
In the limiting case $\xi_{\rm cl}\to0$--namely $N_{\rm cl}\to\infty$ and $\eta_{\rm cl}\to0$-- we find $\tau_{\rm cl}\to0$ and $\tau_{\rm eff}\to\tau_{\rm hom}$, i.e. the homogeneous case.
The case $\xi_{\rm cl}\to0$ corresponds to infinitely many and infinitely compact (and dense, i.e. $C_{\rm cl}\gg1$) clumps.
This resembles the case of a single clump $=$ a single grain, i.e. the homogeneous case. In another limiting case, $\xi_{\rm cl}\to\infty$--that is, $\eta_{\rm cl}\to0$ but $N_{\rm cl}$ is still finite--we find $\tau_{\rm cl}\to\infty$ and $\tau_{\rm eff}\to0$; i.e. there is no dust absorption.
This is a case with few, infinitely compact (and dense) clumps, and radiation escapes easily from the medium.
This is ultimately an inhomogeneous case.
We also consider yet another case, with $\tau_{\rm cl}=\tau_{\rm hom}\xi_{\rm cl}\gg1$ for a finite $\xi_{\rm cl}$.
Recalling that $P_{\rm esc}^{\rm hom}(\tau)\to3/(4\tau)$ when $\tau\gg1$, the effective optical depth becomes $\tau_{\rm eff}\approx\tau_{\rm hom}P_{\rm esc}^{\rm hom}(\tau_{\rm cl})\approx3/(4\xi_{\rm cl})$ when $\tau_{\rm cl}\gg1$.
Therefore, the effective optical depth in a clumpy medium has a maximum value that is determined by the clumpiness of the medium, $\xi_{\rm cl}$.

\subsection{Determining the temperature and mass of dust from observations}

Under the assumption of a modified black-body spectrum for the dust emission (i.e. Kirchhoff's law for an object in thermal equilibrium), we can write the flux density of the observed dust emission as 
\begin{equation}
 F_\nu^{\rm obs}=\frac{1+z}{d_{\rm L}^2} M_{\rm d} \kappa_\nu \left\{B_\nu(T_{\rm d})-B_\nu(T_{\rm CMB})\right\}\,, 
 \label{eq:fnu}
\end{equation}
where $z$ is the redshift and $d_{\rm L}$ is the luminosity distance. 
The CMB term in the parenthesis is the correction term for detection against the CMB in interferometric observations \citep{daCunha13}. 
The dust mass $M_{\rm d}$ is the normalization of the equation and is determined by the observed flux density if the dust temperature $T_{\rm d}$ is given.
If we have multiple data at different wavelengths, the temperature can be determined from them by using an assumed emissivity, $\kappa_\nu$.
However, $T_{\rm d}$ and $M_{\rm d}$ tend to be degenerate, as we shall see in section~\ref{mBBcase} below: a higher $T_{\rm d}$ yields a smaller $M_{\rm d}$. Therefore, we need as many data points as possible to break the degeneracy. 
In particular, a data point at a wavelength on the Wien side--i.e. below the peak wavelength of the flux density--is quite important.

{\it Assuming radiative equilibrium provides an alternative way to break the degeneracy.}
The luminosity absorbed by the dust, $L_*^{\rm abs}$, is a function of the mass of dust $M_{\rm d}$, the source size $R$, the escaping UV luminosity $L_{\rm UV}^{\rm esc}$, and the clumpiness parameter $\xi_{\rm cl}$, and by the dust-grain properties ($\kappa_{\rm UV}$, $\kappa_{\rm IR,0}$, $\nu_0$, and $\beta$). 
The dust temperature in equation~(\ref{TdEq}) can therefore be expressed as 
\begin{equation}
 T_{\rm d}=function(M_{\rm d}, R, L_{\rm UV}^{\rm esc} , \xi_{\rm cl})\,.
 \label{eq:TdMd}
\end{equation}
Thus, in radiative equilibrium
$T_{\rm d}$ and $M_{\rm d}$ are not independent but have a one-to-one connection.
Comparing the flux density from equation~(\ref{eq:fnu}) with observations even in a single band therefore yields $T_{\rm d}$ and $M_{\rm d}$ simultaneously if we know $R$ and $L_{\rm UV}^{\rm esc}$ and assume the dust properties and the clumpiness parameter. 
If there are multiple data points, we may even constrain the clumpiness parameter in the ISM from the dust emission.

\subsection{Dust-grain properties}\label{subsec:dust}

\begin{table*}
  \centering
  \caption{Dust properties.}
  \label{tab:dust}
  \begin{tabular}{lcccl}
    \hline
    & $\kappa_{\rm UV}$$^a$ & $\kappa_{\rm IR,0}$$^b$ & $\beta$ & Remarks/References \\
    & ($10^4$ cm$^2$ g$^{-1}$) & (cm$^2$ g$^{-1}$) & & \\
    \hline
    Fiducial & 5.0 & 30 & 2.0 & \\
    \hline
    Empirical estimates \\
    DustPedia late spirals & --- & $19_{-9}^{+19}$ & (2.0)$^c$ & Converted from the value at 250~$\mu$m; Bianchi et al.~(2019) Fig.~5 right \\
    Milky Way Cirrus & --- & $16\pm3$ & 1.6 & Converted from the value at 250~$\mu$m; 
    Bianchi et al.~(2019) Table~1 \\
    M74 & --- & 28--63 & (2.0)$^c$ & Converted from the value at 500~$\mu$m; 
    Clark et al.~(2019) \\
    M83 & --- & 38--200 & (2.0)$^c$ & Converted from the value at 500~$\mu$m;
    Clark et al.~(2019) \\
    \hline
    Theoretical models \\
    Graphite & 8.16 (4.16) & 52.0 (51.0) & 2.0 & MRN (0.1~$\mu$m)$^d$; Draine \& Lee (1984) \\
    Astronomical silicate & 4.32 (2.43) & 33.3 (32.9) & 2.0 & MRN (0.1~$\mu$m)$^d$; Draine \& Lee (1984), Weingartner \& Draine (2000) \\
    SiC & 7.31 (2.22) & 2.1 (2.1) & 2.0 & MRN (0.1~$\mu$m)$^d$; Laor \& Draine (1993) \\
    Amorphous carbon & 9.91 (5.52) & 54.2 (55.2) & 1.4 & MRN (0.1~$\mu$m)$^d$; Zubko et al.~(1996) \\
    THEMIS & 4.45 & 33.0 & 1.8 & Jones et al.~(2017) CM model; Bianchi et al.~(2019) Table~1 \\
    \hline
    Laboratory measurements & & & & Appendix~A \\
    Crystalline silicates & --- & 0.01--30 & $\sim$ 2--5 & For 4--24 K \\
    Amorphous silicates & --- & 30--260 & $\sim$ 2--3 & For 10--50 K \\
    \hline
  \end{tabular}
  \begin{flushleft}
      $^a$UV mass absorption coefficient. \\
      $^b$Specific emissivity at the wavelength 100 $\mu$m. \\
      $^c$The assumed spectral index. \\
      $^d$Integrated over the MRN size distribution \citep{MRN77}, or for a single size of 0.1~$\mu$m.
  \end{flushleft}
\end{table*}

In the formulation in this paper, the dust properties affect the temperature and mass estimates through the mass absorption coefficient defined by equation~(\ref{MACdef}). 
Specifically, we need to choose the values of $\kappa_{\rm UV}$, $\kappa_{\rm IR,0}$, and $\beta$.
Table~\ref{tab:dust} lists some values for these parameters from the literature.
For $\kappa_{\rm UV}$, we consider theoretical dust models with the standard grain-size distribution for the diffuse ISM in the Milky Way --MRN size distribution-- \citep{MRN77} as well as for some single-sized cases (see Appendix~\ref{sec:MAC1}). 
The assumed UV wavelength range is 0.1--0.3~$\mu$m.
After reviewing these model values, we have adopted $\kappa_{\rm UV}=5.0\times10^4$ cm$^2$ g$^{-1}$ as a fiducial value.
For $\kappa_{\rm IR,0}$ and $\beta$, after considering laboratory measurements, theoretical models, and empirical estimates, we have adopted $\kappa_{\rm IR,0}=30$ cm$^2$ g$^{-1}$ and $\beta=2.0$ as our fiducial set.
We discuss the effects of different values of $\kappa_{\rm UV}$ and $\kappa_{\rm IR,0}$ in section~\ref{sec:MACeffect}.
We also consider cases with $\beta=1$ and 1.5 in section~\ref{mBBcase}.
In the rest of this subsection, we discuss the IR mass absorption coefficient, or emissivity, studies, which is important for the interpretation of IR SED observations.

Many experiments have measured the IR emissivity, $\kappa_{\rm IR}$, of dust-grain analogues in the laboratory.
We briefly review these measurements in Appendix~\ref{sec:MAC2}.
As shown in Table~\ref{tab:labdata}, the experimental emissivity values are distributed over several orders of magnitude.
This is partly because the results are sensitive to the experimental conditions, which are difficult to control.
Although direct comparisons among different experiments may not be meaningful, we find that there are general dependencies on crystallinity and temperature.
For example, the values of $\kappa_{\rm IR}$ for amorphous materials are about an order of magnitude larger than those of crystalline materials.
Also, for lower temperatures, $\kappa_{\rm IR}$ tends to be lower.
The index $\beta$ also depends on crystallinity: 
amorphous materials tend to have smaller values of $\beta$ than crystalline materials,
and $\beta$ tends to be larger at lower temperatures.
Given the large variation in the measurements and the lack of knowledge about the composition and crystallinity of the actual dust grains, especially in high-$z$ galaxies, it is difficult at present to use the laboratory measurements directly.
However, in the future it may be worth considering the temperature dependencies in fitting IR SEDs.

Next, we consider the use of theoretical dust models.
Several models successfully reproduce the extinction curves and the IR SEDs in the Galaxy and the Magellanic Clouds (e.g. \citealt{DL84,Zubko96,WD01,DraineLi07,Jones17}).
These models are based on the complex refractive index of some kind of dust material and are calculated by using Mie theory, assuming compact spherical grains with a given size distribution.
In the middle part of Table~\ref{tab:dust}, we list the IR emissivity values of four theoretical models taken from the listed references.
These values are typically 30--50 cm$^{2}$ g$^{-1}$ at the wavelength of $100 \micron$, and they are broadly consistent with the laboratory data.
The index $\beta$ is expected to be around 2.
However, the actual FIR SEDs of galaxies often give a smaller value of $\beta$, around 1.5.
The cause of this discrepancy is unknown, but the dependence of the temperature distribution on grain size and environment may be one reason.

Recently, there have been attempts to measure the IR emissivities empirically in the Galaxy and nearby galaxies \citep{Clark16,Clark19,Bianchi19}.
The method is based on the observed ratio of the surface brightness at FIR and submillimetre wavelengths to the surface density of the gas mass in conjunction with measurements of metallicity and elemental depletion (i.e. the dust-to-metal ratio).
The emissivity values at 100~$\mu$m--converted from those obtained in these papers by assuming $\beta=2$--are listed in the top part of Table~\ref{tab:dust}.
Although the $\kappa_{\rm IR}$ values show significant variations in galaxy disks, a typical value is 20--30 cm$^2$ g$^{-1}$ at $100 \micron$, which is somewhat smaller than for the theoretical models.
This empirical estimate may provide a good {\it ansatz} for the emissivity averaged over a galaxy-wide scale, including the effects of variations in the temperature, the size distributions, and the composition of the dust.

\subsection{Overall behaviour}\label{Overall}

\begin{figure}
 \begin{center}
  \includegraphics[width=6cm]{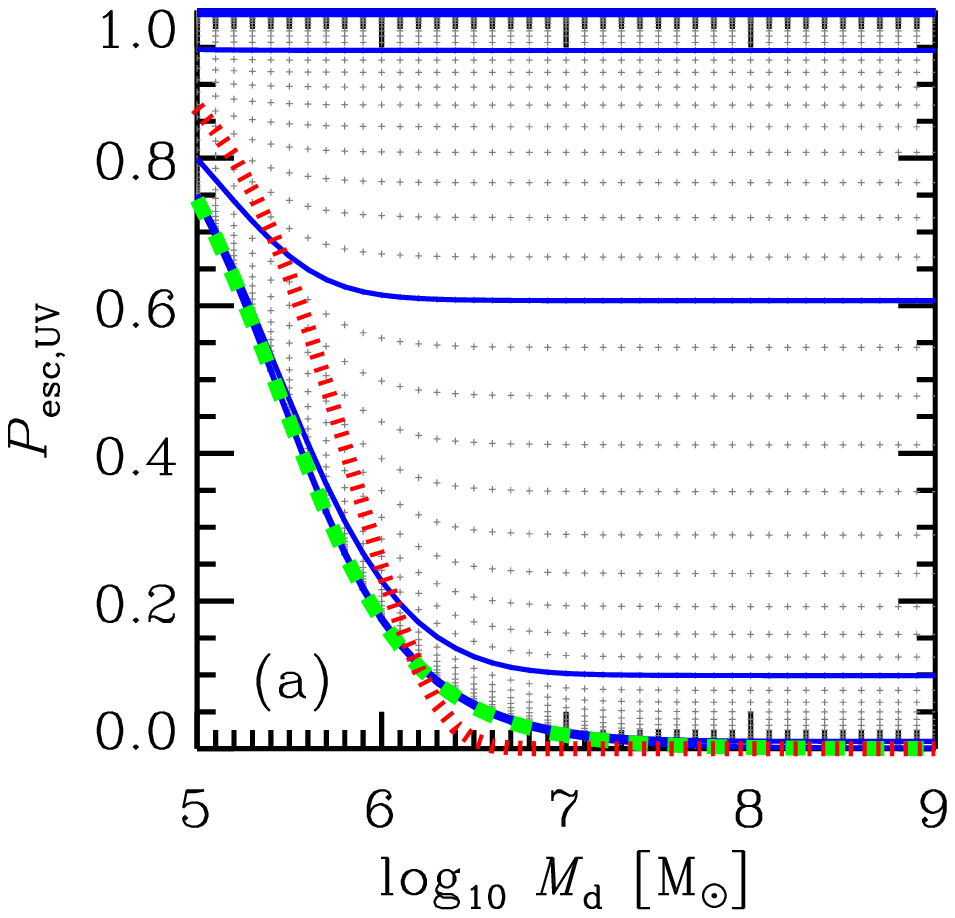}
  \includegraphics[width=6cm]{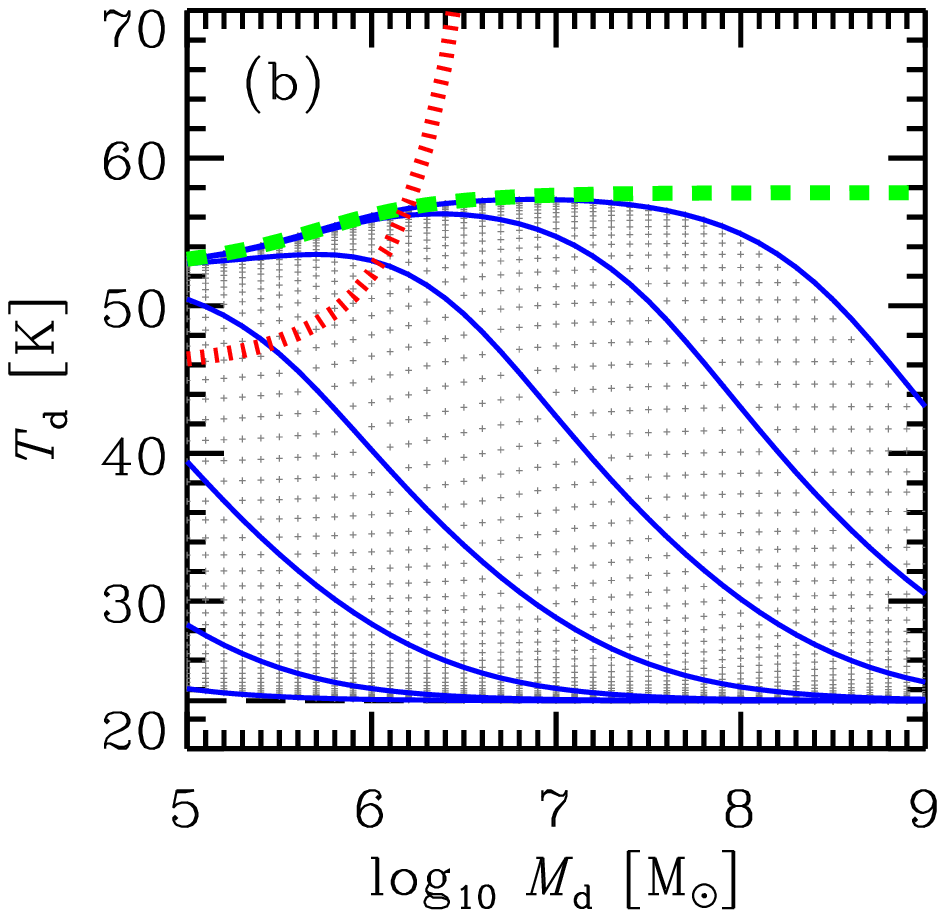}
  \includegraphics[width=6cm]{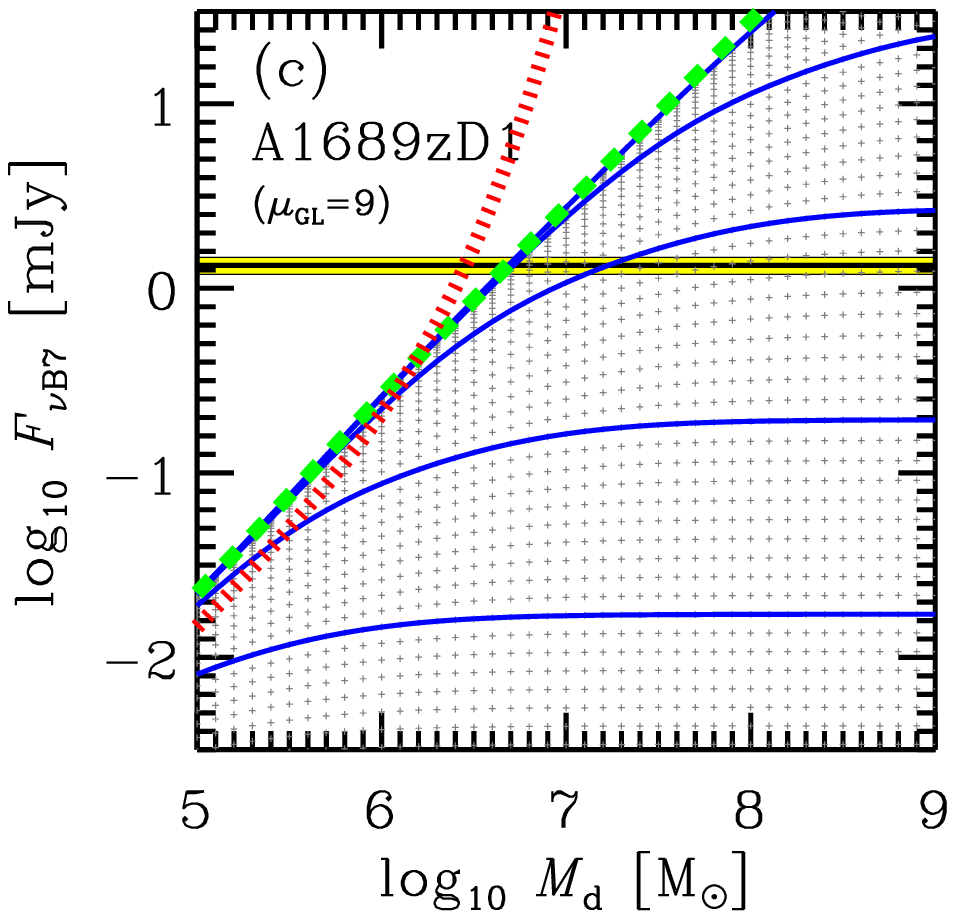}
 \end{center}
 \caption{(a) The UV escape probability $P_{\rm esc,UV}$, (b) dust temperature $T_{\rm d}$, and (c) Band~7 flux density $F_{\nu_{\rm B7}}$, as functions of the dust mass $M_{\rm d}$ for the A1689zD1 parameters listed in Table~1 (i.e. representative values of $L_{\rm UV}^{\rm esc}$ and $R$). The red dotted and green dashed lines represent the spherical shell and homogeneous sphere cases, respectively. The blue solid lines are clumpy cases, with the clumpiness parameters $\log_{10}\xi_{\rm cl}=-3$, $-2$, $-1$, $0$, $1$, $2$, and $3$ from bottom to top in (a), from top-right to bottom-left in (b), and from top to bottom in (c). The small grey plus-signs show sequences with other values of $\log_{10}\xi_{\rm cl}$. In panel~(b), the black long-dashed horizontal line indicates the cosmic microwave-background temperature at the object redshift (23.25~K), which sets the lower limit to $T_{\rm d}$. In panel (c), the horizontal yellow shading shows the observed Band~7 flux density and its $\pm1\sigma$ range. These calculations employed the fiducial dust emissivity given in Table~\ref{tab:dust} and the lensing magnification $\mu_{\rm GL}=9$.}
 \label{fig:1Dplot}
\end{figure}

\begin{figure}
  \begin{center}
    \includegraphics[width=6.5cm]{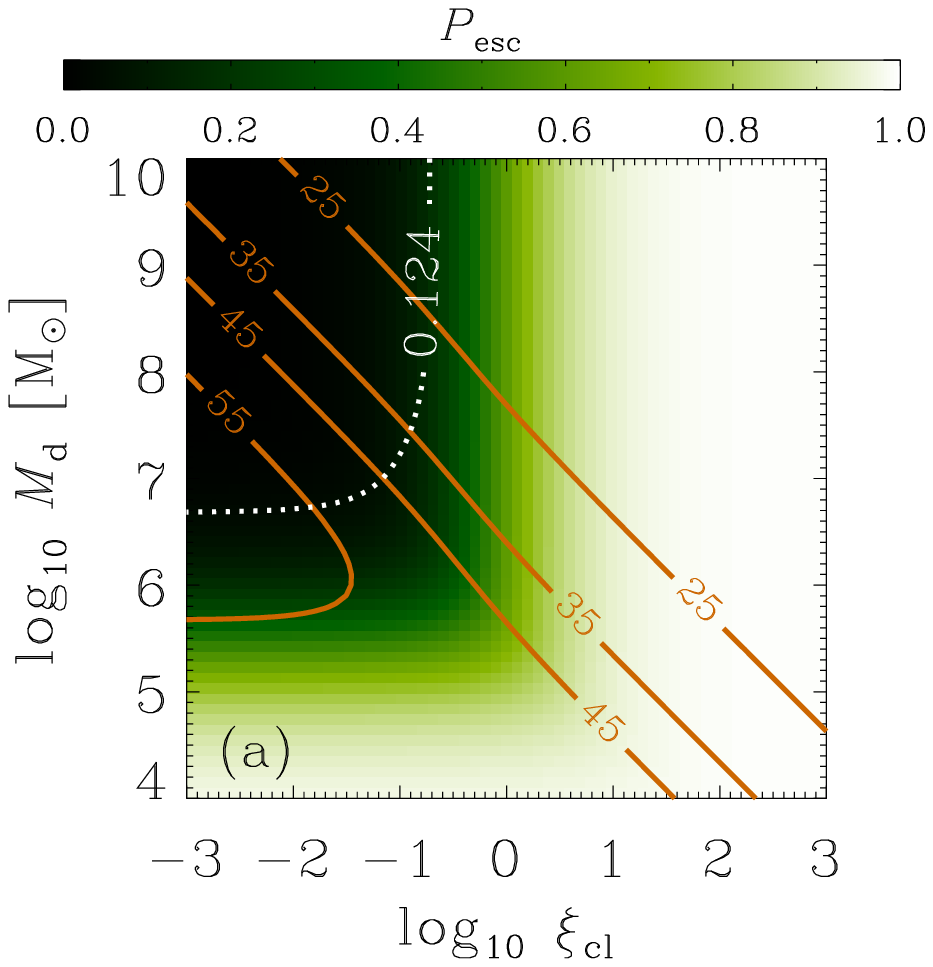}
    \includegraphics[width=6.5cm]{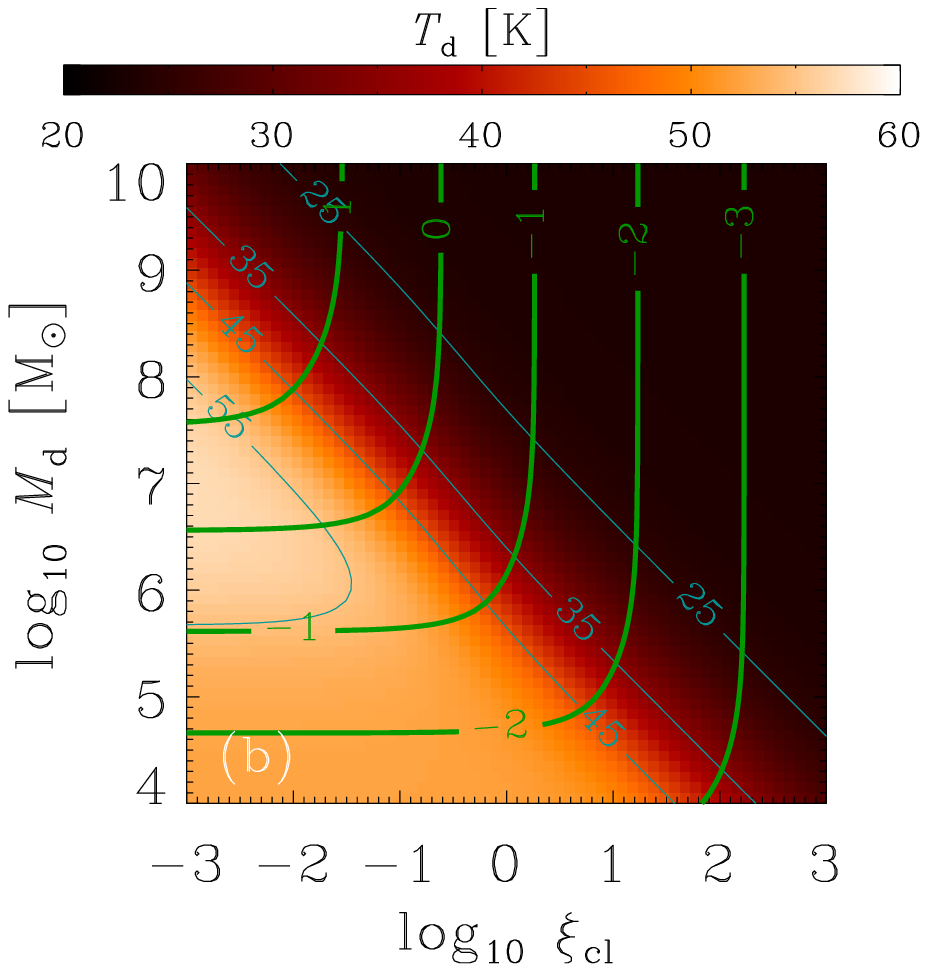}
    \includegraphics[width=6.5cm]{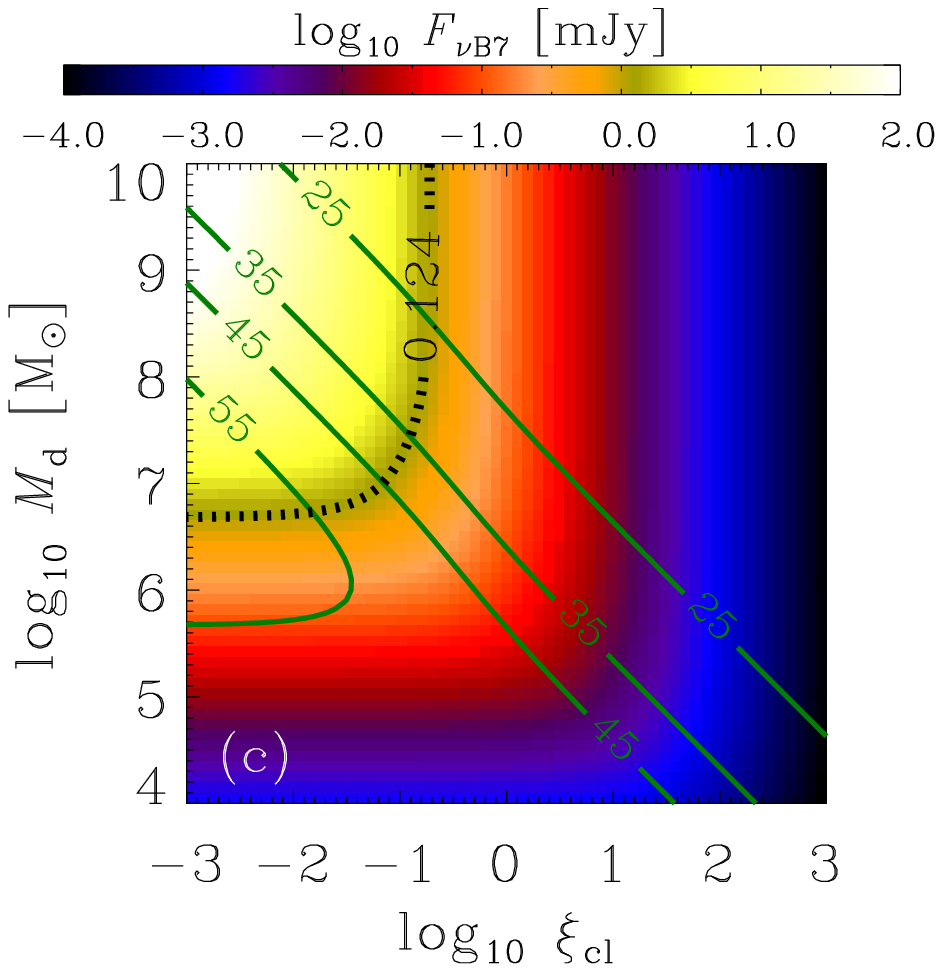}
  \end{center}
  \caption{(a) The UV escape probability $P_{\rm esc,UV}$, (b) dust temperature $T_{\rm d}$, and (c) Band~7 flux density $F_{\nu_{\rm B7}}$ in the $(M_{\rm d}, \xi_{\rm cl})$ plane for the A1689zD1 parameters listed in Table~1 (i.e. representative values of $L_{\rm UV}^{\rm esc}$ and $R$). All panels show the contours of $T_{\rm d}$. In panels (a) and (c), the dotted line shows the $(M_{\rm d}, \xi_{\rm cl})$ combinations that give the observed Band~7 flux density, $\log_{10}(F_{\nu_{\rm B7}}/{\rm mJy})=0.124$. The green contours in panel (b) show the Band~7 flux density in mJy in the common logarithmic scale. These calculations employed the fiducial dust emissivity given in Table~\ref{tab:dust} and the lensing magnification $\mu_{\rm GL}=9$.}
  \label{fig:2Dplot}
\end{figure}

Here we look into the overall behaviour of the radiative-equilibrium models for the three geometries we consider.
Given the object size $R$ and the escaping UV luminosity $L_{\rm UV}^{\rm esc}$ (and $\xi_{\rm cl}$ for the clumpy geometry), we can solve the radiative-equilibrium models.
Figure~\ref{fig:1Dplot} (a) shows the UV escape probability $P_{\rm esc,UV}$ as a function of $M_{\rm d}$.
For the shell case (the red dotted line), $P_{\rm esc,UV}$ decreases exponentially as $M_{\rm d}$ (or equivalently the system optical depth) increases.
The homogeneous-sphere case (the green dashed line) exhibits a slower decline than the shell case: $P_{\rm esc,UV}\propto 1/M_{\rm d}$ for $M_{\rm d}\to\infty$, as expected from equation~(\ref{eq:P_hom}).
For the clumpy cases (the blue solid lines and grey plus-signs), $P_{\rm esc,UV}$ follows the trend of the homogeneous case when the clumpiness parameter $\xi_{\rm cl}$ and the single-clump optical depth $\tau_{\rm cl}$ are small enough, but it deviates from the homogeneous case and approaches the constant value $P_{\rm esc}^{\rm hom}(3/4\xi_{\rm cl})$ when $\tau_{\rm cl}$ becomes large.
These behaviours can be well understood from the discussion in the last paragraph of section~\ref{sec:clumpy}.

Panels~(b) and (c) of Figure~\ref{fig:1Dplot} show $T_{\rm d}$ and the flux density in Band~7, $F_{\rm \nu B7}$, as functions of $M_{\rm d}$.
As found from equation~(\ref{TdEq}), $T_{\rm d}$ is mainly determined by the ratio $L_*^{\rm abs}/M_{\rm d}$.
The absorbed stellar-radiation energy, $L_*^{\rm abs}$, can be approximated by $L_*^{\rm abs}\propto1/P_{\rm esc,UV}$ when $P_{\rm esc,UV}$ is small (see eq.~[\ref{eq:Labs}]).
For the shell case, we find $L_*^{\rm abs}\propto e^{M_{\rm d}}$ according to equation~(\ref{eq:P_shell}).
Therefore, $T_{\rm d}$ increases rapidly as $M_{\rm d}$ increases, as shown by the red dotted line in panel~(b).
The corresponding flux $F_{\rm \nu B7}$ also increases rapidly, as shown in panel~(c).
For the homogeneous case, $L_*^{\rm abs}\propto M_{\rm d}$ from equation~(\ref{eq:P_hom}) when $\tau_{\rm hom}$ is large.
Thus, $T_{\rm d}$ reaches a constant value when $M_{\rm d}$ is large enough, and $F_{\rm \nu B7}$ is linearly proportional to $M_{\rm d}$, as shown by the green dashed lines in panels~(b) and (c), respectively.

In clumpy media, $L_*^{\rm abs}$ follows the homogeneous case for small $M_{\rm d}$, but it approaches a constant value determined by $\xi_{\rm cl}$ when $M_{\rm d}$ is sufficiently large, as does $P_{\rm esc,UV}$ (see panel~[a]).
Therefore, $T_{\rm d}$ decreases as $M_{\rm d}$ increases after deviating from the homogeneous case, as shown by the blue solid lines in panel~(b).
The corresponding flux $F_{\rm \nu B7}$ reaches a constant value, as shown in panel~(c).
This means that an infinitely large amount of dust can exist in clumpy media while a constant continuum flux density is maintained, which is unphysical.
This artefact is caused by the unlimited dust density in clumps in our formulation, allowing us to hide an infinitely large amount of dust in clumps that do not produce any UV absorption.
In reality, there must be an upper limit to the dust density in clumps, which is determined by-- at least--the material density of the dust grains and the stability of the clumps against their own self-gravity.
We also neglected self-absorption of the dust continuum, which becomes significant when $M_{\rm d}$ is sufficiently large.

Figure~\ref{fig:2Dplot} shows the distributions of $P_{\rm esc,UV}$, $T_{\rm d}$, and $F_{\rm \nu B7}$ in the ($M_{\rm d},\xi_{\rm cl}$) plane for the clumpy models.
As can be seen from Figure~\ref{fig:1Dplot} (c), the two parameters $M_{\rm d}$ and $\xi_{\rm cl}$ in the clumpy models are degenerate when only a single IR observation is available, while a unique solution for $M_{\rm d}$ can be found in the shell and homogeneous cases.
This is clearly shown in the two-dimensional plots.
Fortunately, the contours (or dependence) of $T_{\rm d}$ and $F_{\rm \nu B7}$ are not parallel.
If $T_{\rm d}$ is constrained by the IR SED shape, then, we can determine $M_{\rm d}$ and $\xi_{\rm cl}$ simultaneously.
This implies
{\it the interesting possibility of discussing the clumpiness of the ISM by using spatially unresolved IR SEDs of galaxies.}

\section{Dust temperature and mass of A1689zD1}

We are now ready to apply our algorithm to determine the dust temperature $T_{\rm d}$ and mass $M_{\rm d}$ in the high-$z$ galaxy A1689zD1.
To compare $T_{\rm d}$ and $M_{\rm d}$ from our method with those from the usual method, we first present modified black-body fits in section~\ref{mBBcase}.
We then present the shell- and homogeneous-geometry cases in section~\ref{sec:shelhomo} and the clumpy cases in section~\ref{sec:clpresult}.
In the last section~\ref{sec:MACeffect}, we examine the effects of the mass absorption coefficient $\kappa$ on the determination of $T_{\rm d}$ and $M_{\rm d}$.

\begin{table*}
  \centering
  \caption{A summary of the derived dust temperatures and masses of A1689zD1 for modified black-body cases and for a shell, homogeneous sphere, and clumpy medium in radiative equilibrium.}
  \label{tab:fitresults}
  \begin{tabular}{lccccc}
    \hline
    Cases & $T_{\rm d}$ [K] (68\% range) & $\log_{10}M_{\rm d}$ [M$_\odot$] (68\% range) 
    & $\log_{10}\xi_{\rm cl}$ (68\% range) 
    & $\log_{10}L_{\rm IR}$ [L$_\odot$] (68\% range) 
    & $SFR_{\rm IR}$ [M$_\odot$ yr$^{-1}$] \\
    \hline
    \multicolumn{5}{l}{Modified BB, all 3 bands} \\
    $\beta=1$ & 66.0 (50.5--99.8) & 6.51 (6.12--6.82) & --- & 11.58 (11.30--12.10) & 65 \\
    $\beta=1.5$ & 47.9 (39.1--62.9) & 6.91 (6.57--7.21) & --- & 11.38 (11.19--11.69) & 41 \\
    $\beta=2$ & 37.7 (31.6--46.5) & 7.30 (6.96--7.66) & --- & 11.26 (11.15--11.48) & 31 \\
    \hline
    \multicolumn{5}{l}{$\beta=2.0$, all 3 bands} \\
    Shell & 70.3 (65.2--76.7) & 6.45 (6.36--6.53) & --- & 12.04 (11.91--12.19) & 190 \\
    Homogeneous & 57.3 (54.7--60.3) & 6.69 (6.61--6.75) & --- & 11.74 (11.67--11.82) & 93 \\
    Clumpy & 37.8 (31.5--46.7) & 7.29 (6.96--7.67) & $-0.97$ ($-1.24$--$-0.82$) & 11.24 (11.08--11.48) & 30 \\
    \hline
    \multicolumn{5}{l}{$\beta=2.0$, only Band~6} \\
    Shell & 79.6 (71.4--89.6) & 6.54 (6.45--6.62) & --- & 12.45 (12.21--12.72) & 480 \\
    Homogeneous & 57.4 (54.8--60.3) & 6.82 (6.72--6.90) & --- & 11.88 (11.76--11.99) & 130 \\
    \hline
    \multicolumn{5}{l}{$\beta=2.0$, only Band~7} \\
    Shell & 70.5 (65.1--77.1) & 6.45 (6.36--6.53) & --- & 12.05 (11.90--12.21) & 190 \\
    Homogeneous & 57.3 (54.7--60.2) & 6.68 (6.61--6.75) & --- & 11.74 (11.65--11.82) & 93 \\
    \hline
    \multicolumn{5}{l}{$\beta=2.0$, only Band~8} \\
    Shell & 67.0 (61.6--73.1) & 6.41 (6.30--6.50) & --- & 11.87 (11.67--12.05) & 130 \\
    Homogeneous & 57.3 (54.6--60.2) & 6.61 (6.49--6.71) & --- & 11.66 (11.54--11.77) & 78 \\
    \hline
  \end{tabular}
\end{table*}

\subsection{Modified black-body cases}\label{mBBcase}

\begin{figure}
  \begin{center}
    \includegraphics[width=8cm]{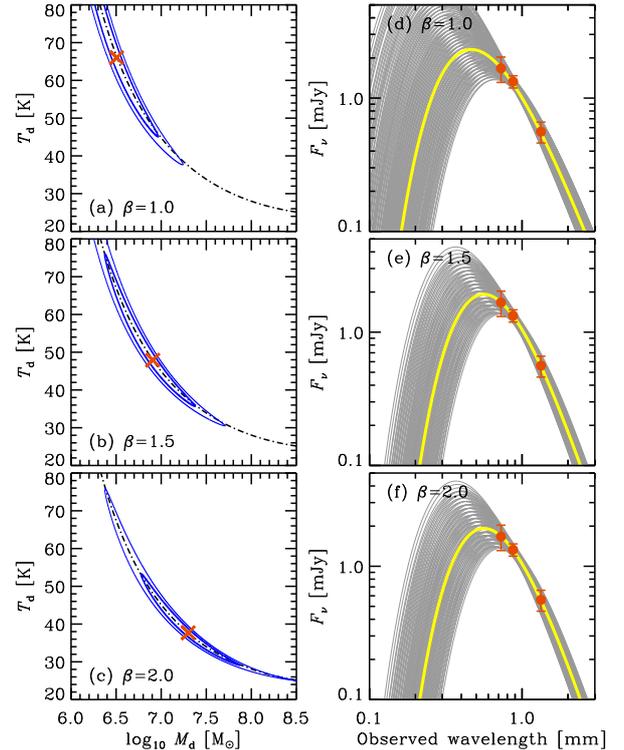}
  \end{center}
  \caption{Results of modified black-body fits to the IR SED of A1689zD1.
  (a--c) The best-fit solutions (crosses) and the central 68\% (thick solid lines) and 95\% (thin solid lines) areas in the $T_{\rm d}$--$M_{\rm d}$ plane for the emissivity indices $\beta=1.0$, 1.5, and 2.0. The dot-dashed lines show the positions of ($T_{\rm d}$,$M_{\rm d}$) that provide the observed Band~7 flux density.
  (d--f) SEDs for the best-fit solutions (thick yellow lines) and for those in the central 68\% areas of the ($T_{\rm d}$,$M_{\rm d}$) plane (thin gray lines) for $\beta=1.0$, 1.5, and 2.0.}
  \label{fig:mBBfit}
\end{figure}

First, we obtained a modified black-body fit to the three ALMA observations of the IR SED of A1689zD1 listed in Table~\ref{tab:obsdata}.
We have adopted a two-dimensional $\chi^2$ minimization method to find the best-fit solution in the $(T_{\rm d}, M_{\rm d})$ plane. 
Specifically, we have searched for the point in the $(T_{\rm d}, M_{\rm d})$ plane that yields the minimum value of $\chi^2=\sum_{i=1}^3 (F_{\nu_i}^{\rm exp}-F_{\nu_i}^{\rm obs})^2/\sigma_{\nu_i}^2$, where $F_{\nu_i}^{\rm exp}$ is the flux density at frequency $\nu_i$ expected from equation~(\ref{eq:fnu}), $F_{\nu_i}^{\rm obs}$ is the observed flux density, and $\sigma_{\nu_i}$ is its uncertainty, and where both $T_{\rm d}$ and $M_{\rm d}$ are changed simultaneously.
Here, we have assumed the fiducial dust emissivity given in Table~\ref{tab:dust}.
We also consider cases with the indices $\beta=1.0$ and 1.5.

Figure~\ref{fig:mBBfit} shows the modified black-body fits, and the top part of Table~\ref{tab:fitresults} gives a summary of the best-fit sets of $T_{\rm d}$ and $M_{\rm d}$ and their 68\% ranges.
Since we may not yet have constrained the peak of the IR SED, $T_{\rm d}$ and its associated $M_{\rm d}$ still have large uncertainties.\footnote{If we force the spectrum to pass through the flux density measured in a given band, the uncertainties in both $T_{\rm d}$ and $M_{\rm d}$ become much smaller (e.g. \citealt{Hashimoto19}).}
The emissivity $\kappa_{\rm IR,0}$ affects the estimated $M_{\rm d}$ linearly for fixed $T_{\rm d}$, whereas the effect of the emissivity index $\beta$ is non-linear.
From Figure~\ref{fig:mBBfit}, we find that a smaller $\beta$ yields a higher $T_{\rm d}$ and a smaller $M_{\rm d}$.
For a higher $T_{\rm d}$, the IR SED in the observed Bands~8, 7, and 6 approaches the Rayleigh-Jeans limit, where a larger $\beta$ gives a steeper SED slope.
For a lower $T_{\rm d}$, on the other hand, Band~8 comes closer to the SED peak, and the slope among the three bands becomes shallower.
For a given $\beta$, $T_{\rm d}$ is determined by the interplay between these two opposite effects.
Since a smaller $\beta$ makes the SED slope in the Rayleigh-Jeans limit shallower, a higher $T_{\rm d}$ is favoured by the observed data.
For a larger $\beta$, on the other hand, the Rayleigh-Jeans slope is already steep, and consequently a lower $T_{\rm d}$ is favoured.

In Table~\ref{tab:fitresults}, top part, the dust IR luminosities for the best-fit cases, their corresponding 68\% ranges, and the SFRs are also given. 
We estimated these IR-based SFRs by using the conversion formula of \cite{Kennicutt98}. 
Compared to the UV-based SFR given in Table~\ref{tab:obsdata}, these IR-based SFRs are a factor of 5--10 larger in the standard modified black-body fit.
Therefore, the observed UV radiation traces only a 10--20\% of the total SFR in the high-$z$ galaxy, A1689zD1, and the dominant part of the SFR is obscured by dust.

\subsection{Radiative equilibrium cases in the shell and homogeneous geometries}\label{sec:shelhomo}

Here we obtain fits by adopting radiative equilibrium in the shell and homogeneous spherical geometries.
We first show the results using data from all three bands and then the results obtained by using single-band data.
The resulting values are summarised in Table~\ref{tab:fitresults}.

\subsubsection{Multi-band fit}

\begin{figure}
 \begin{center}
   \includegraphics[width=6cm]{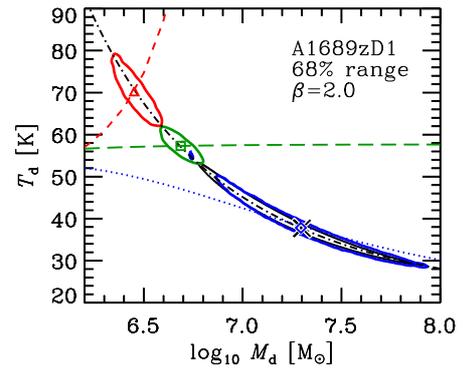} 
 \end{center}
 \caption{The best-fit solutions (symbols) and central 68\% areas (solid lines) in the $T_{\rm d}$--$M_{\rm d}$ plane for four models, with the emissivity index $\beta=2.0$. The black cross and line are the same modified black-body fit as in Fig.~\ref{fig:mBBfit} (c). The red triangle and line are the spherical thin-shell case, the green square and line are the homogeneous-sphere case, and the blue diamond and line are the clumpy-sphere case. The black dot-dashed line indicates the $T_{\rm d}$--$M_{\rm d}$ relation expected from the observed Band~7 flux density (as in Fig.~\ref{fig:mBBfit} [c]). The red short-dashed, green long-dashed, and blue dotted lines are the tracks expected from radiative equilibrium for the shell, homogeneous, and clumpy geometries, respectively. For the dotted line, we adopted the parameter $\log_{10}\xi_{\rm cl}=-1.0$ as the best-fit solution.}
 \label{fig:TdlogMd}
\end{figure}

As we saw in section 3.3, there is a one-to-one connection between $T_{\rm d}$ and $M_{\rm d}$ in equation~(\ref{eq:TdMd}) in radiative equilibrium.
We have therefore performed a one-dimensional $\chi^2$ minimization to find the best-fit solution for $M_{\rm d}$ that simultaneously gives the best-fit $T_{\rm d}$.
We estimated the uncertainties in the best-fit $M_{\rm d}$ and the corresponding $T_{\rm d}$ by using a Monte Carlo method to take into account the observational uncertainties in the source radius $R$ as well as in the flux densities.
Specifically, we varied $R$ and the flux densities in each realization, assuming Gaussian distributions with the observed $1\sigma$ uncertainties as the standard deviations, and searched for the `best-fit' $M_{\rm d}$ (and corresponding $T_{\rm d}$) for that set of data.
We repeated this procedure $>10,000$ times.

Figure~\ref{fig:TdlogMd} shows the fitting results for radiative equilibrium.
The red and green dashed lines are the values of $T_{\rm d}$ as a function of $M_{\rm d}$ expected from radiative equilibrium in the thin spherical shell and the homogeneous sphere, respectively.
The slopes of these lines are very different from the black dot-dashed line obtained from equation~(\ref{eq:fnu}) with the Band~7 flux density.
Therefore, radiative equilibrium breaks the $T_{\rm d}$--$M_{\rm d}$ degeneracy; in other words, it determines $T_{\rm d}$.
As a result, the central 68\% areas (shown in red and green) in the $(T_{\rm d}, M_{\rm d})$ plane become smaller than that obtained from the modified black-body fit (shown in black).
The uncertainties (i.e. the extensions of the areas) along the $T_{\rm d}$--$M_{\rm d}$ degeneracy line (the black dot-dashed line) are determined by the uncertainty in the source size $R$, while those perpendicular to the degeneracy line are determined by the flux-density uncertainties.
This is because a smaller $R$ gives a larger $\tau$, smaller $P_{\rm esc}$, larger $L_*^{\rm abs}$, and eventually higher $T_{\rm d}$ for the system, and {\it vice versa}.

The obtained best-fit temperatures and masses are summarised in the second part of Table~\ref{tab:fitresults}.
The IR luminosities and corresponding IR-based SFRs are also listed there.
In the shell case, the dust temperature becomes as high as 70~K, and therefore, the IR luminosity exceeds $10^{12}$ L$_\odot$; the galaxy is classified as a Ultra-Luminous Infrared Galaxy (ULIRG).
The corresponding IR-based SFR is as high as $\sim200$ M$_\odot$.
In the homogeneous sphere case, the dust temperature is slightly lower than that of the shell case but still about 60~K.
The IR-based SFR is as high as $\sim100$ M$_\odot$.
Therefore, the UV-traced SFR is only 3--6\% of the total one in these cases.
Although such a situation may be a true feature of the high-$z$ galaxy, A1689zD1, it could also indicate the invalidity of the assumed simple geometries.

\subsubsection{Single-band fit}

\begin{figure}
  \begin{center}
    \includegraphics[width=6cm]{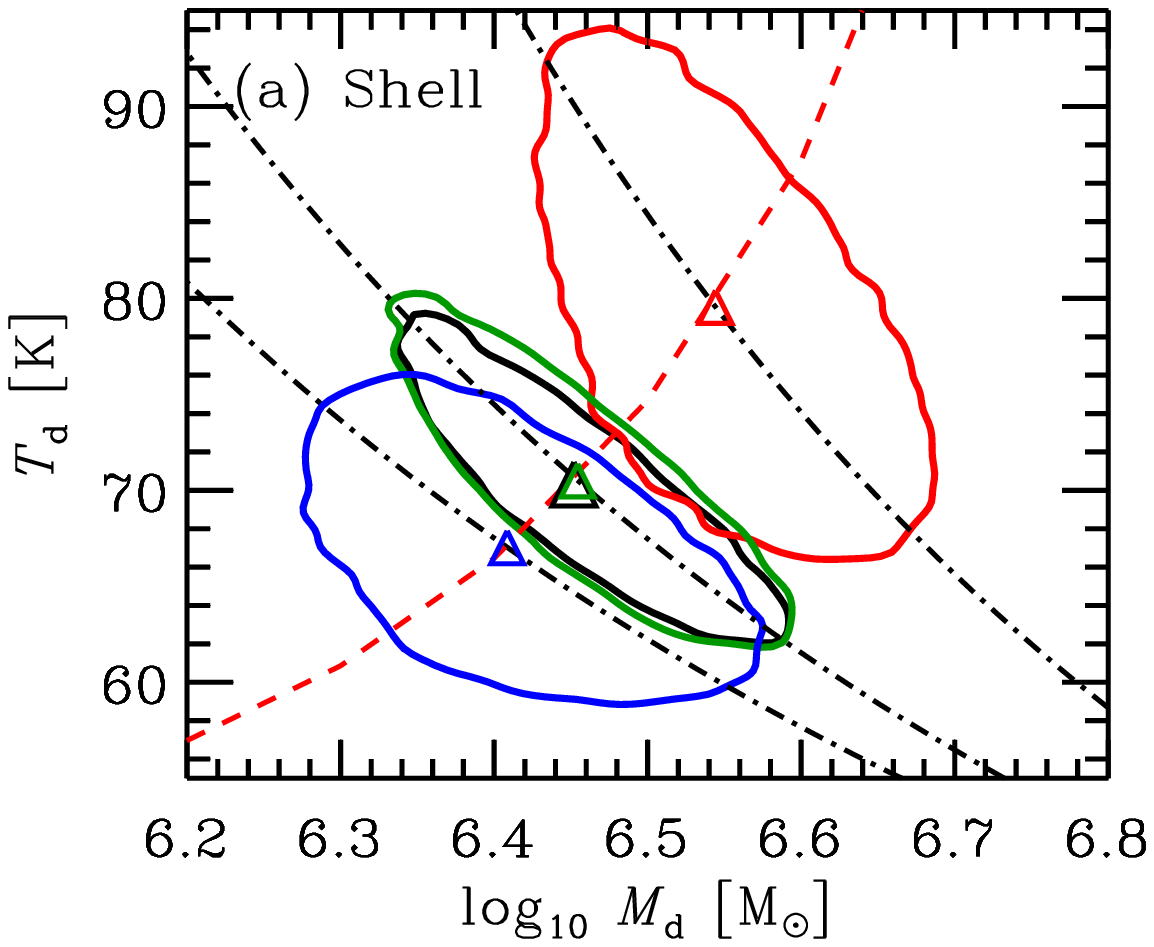}
    \includegraphics[width=6cm]{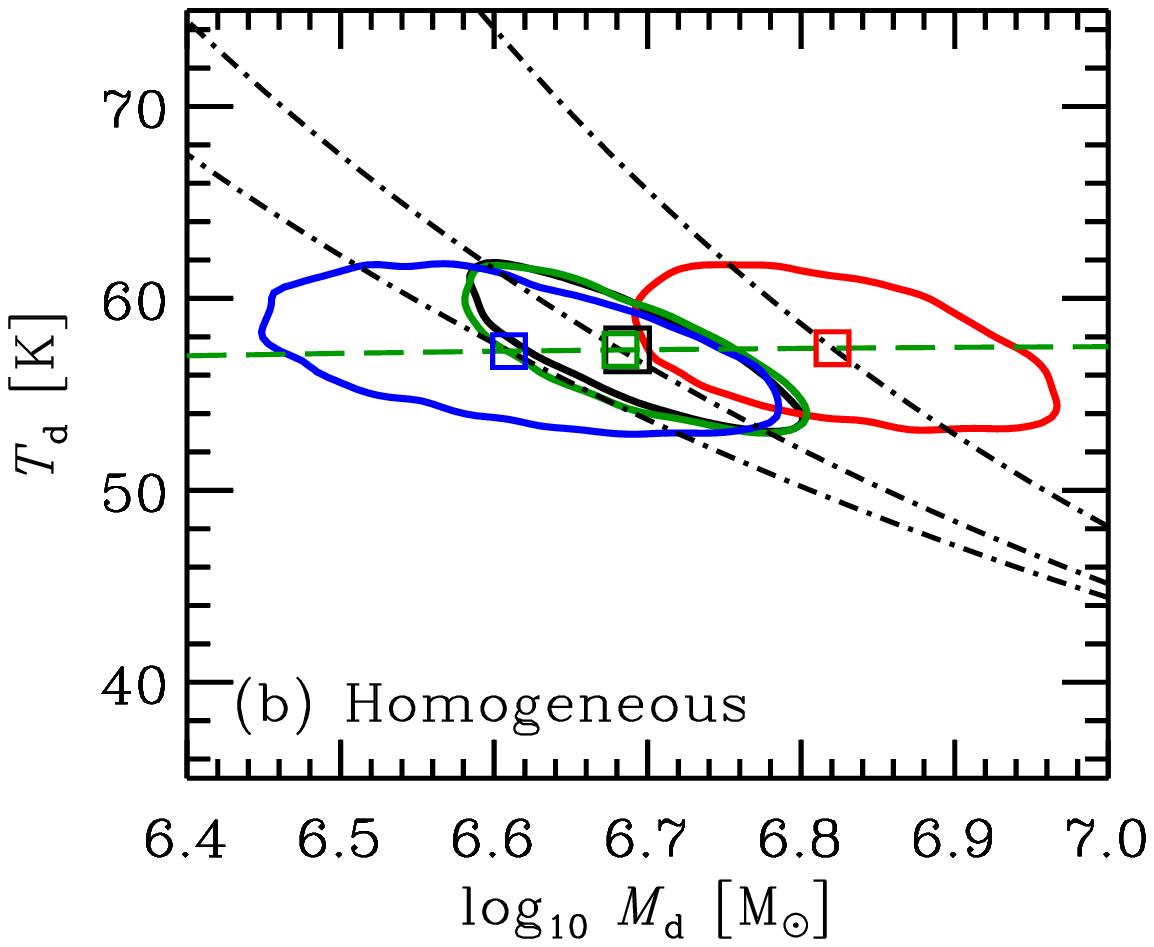}
  \end{center}
  \caption{The same as Fig.~\ref{fig:TdlogMd}, but for cases using data from only a single band for the shell and homogeneous geometries. The red, green, and blue symbols and areas show the Band~6, 7, and 8 cases, respectively, while the black symbol and area (almost overlapping with the green ones) represent the results obtained using all three bands. The black dot-dashed lines are the $T_{\rm d}$--$M_{\rm d}$ relations expected from the observed flux densities in the three bands.}
  \label{fig:TdMd1band}
\end{figure}

Radiative equilibrium gives $T_{\rm d}$ as a function of $M_{\rm d}$ in the shell and homogeneous geometries (see eq.~\ref{eq:TdMd}).
Even a single-band flux density determines $M_{\rm d}$ and the corresponding $T_{\rm d}$ simultaneously.
Figure~\ref{fig:TdMd1band} shows such best-fit ($T_{\rm d}, M_{\rm d}$) solutions using each single-band flux density for the two geometries.
The solutions occur at the intersections of the radiative-equilibrium lines (dashed) and the $T_{\rm d}$--$M_{\rm d}$ degeneracy lines (dot-dashed) given by the flux density in each band.
The Band~7 data give the narrowest 68\% areas, which are slightly larger than but very similar to those obtained by fitting all three band for this object.
The Band~6 and 8 solutions are different from that for Band~7, but they are located in and around the Band~7 68\% area.
On the other hand, the Band~7 (or multi-band) solution occurs within the 68\% areas of Band~8 (blue) but slightly out of the area of Band~6 (red).
Given the current uncertainties in size and flux density, these three single-band solutions are reasonably consistent with the multi-band solution.
The obtained values are summarised in the bottom half of Table~\ref{tab:fitresults}.

\subsection{Radiative-equilibrium case in the clumpy geometry}\label{sec:clpresult}

\begin{figure}
 \begin{center}
   \includegraphics[width=6cm]{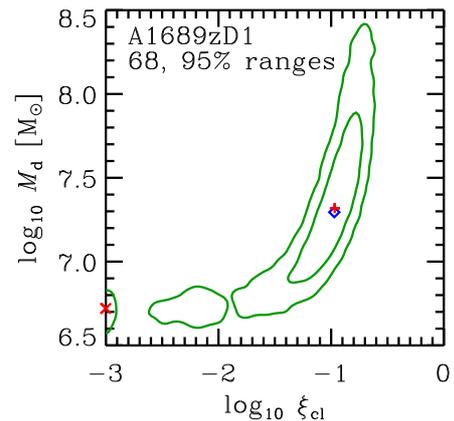}
 \end{center}
 \caption{The two-dimensional distribution of the `best-fit' solutions from 10,000 Monte Carlo runs with varied observational data in the clumpy-geometry fitting. From inside to outside, the contours enclose 68\% and 95\% of the solutions. The highest and second highest peaks of the density of solutions are shown by the plus-sign and cross, respectively. The location of the best-fit solution for the actual observational data is shown by the diamond.}
 \label{fig:Mdclp}
\end{figure}

In the clumpy geometry, we have an additional parameter that quantifies the clumpiness, $\xi_{\rm cl}$.
Therefore we need at least two data points in the IR SED to obtain a robust fit even in radiative equilibrium.
Fortunately, there are three data points for A1689zD1.
We have accordingly performed a two-dimensional $\chi^2$ minimization in the $(M_{\rm d}, \xi_{\rm cl})$ plane.
We again estimated the fitting uncertainties from Monte Carlo simulations, as for the shell and homogeneous cases.
The values obtained are listed in the second part of Table~\ref{tab:fitresults}.
 
In Figure~\ref{fig:TdlogMd}, the best-fit solution and its uncertainty are shown by the blue symbol and line.
Due to the additional parameter $\xi_{\rm cl}$, the best-fit solution in the $(T_{\rm d}, M_{\rm d})$ plane is found to occur at essentially the same position as for the modified black-body case.
We have also found the best-fit value of $\log_{10}\xi_{\rm cl}\simeq-1$.
With this value of $\xi_{\rm cl}$, radiative equilibrium in the clumpy geometry gives the $T_{\rm d}$--$M_{\rm d}$ relation shown by the blue dotted line in Figure~\ref{fig:TdlogMd}.
The uncertainty in the $(T_{\rm d}, M_{\rm d})$ plane is slightly smaller than that of the modified black-body case, but it is larger than for the shell and homogeneous cases, probably because of the two-parameter fit for the clumpy case.
Note that the shell and homogeneous cases have only the single fitting parameter $M_{\rm d}$, while $T_{\rm d}$ is determined by radiative equilibrium, as shown by the red and green dashed lines in Figure~\ref{fig:TdlogMd}.

Figure~\ref{fig:Mdclp} shows the distribution of the best-fit solutions from Monte Carlo runs in the $(M_{\rm d}, \xi_{\rm cl})$ plane.
The distribution is clearly bimodal. 
The first peak in the distribution (the plus-sign) occurs at a position very close to the best-fit solution for the actual data (diamond).
The second peak (the cross) is found on the axis where $\log_{10}\xi_{\rm cl}=-3$, which is the lowest limit of $\xi_{\rm cl}$ in our calculations.
This solution is essentially the same as for the homogeneous case, i.e. the limiting case of $\xi_{\rm cl}\to0$ (see the discussion in section~\ref{sec:clumpy}).

\subsection{Effect of mass absorption coefficients}\label{sec:MACeffect}

\begin{figure}
 \begin{center}
   \includegraphics[width=6cm]{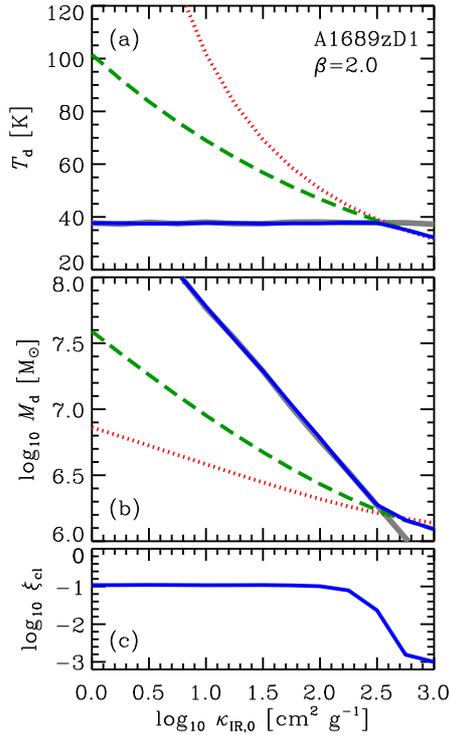}
 \end{center}
 \caption{The effects of the IR mass absorption coefficient $\kappa_{\rm IR,0}$ at the wavelength 100 \micron, assuming the index $\beta=2.0$. (a) Dust temperature, (b) dust mass, and (c) the clumpiness parameter $\xi_{\rm cl}$. The dotted, dashed, and solid lines correspond to the shell, homogeneous-sphere, and clumpy-sphere geometries, respectively. The modified black-body fit is shown by the grey solid line, which almost overlaps with the clumpy case when $\log_{10}\xi_{\rm cl}\simeq-1$. The case $\log_{10}\xi_{\rm cl}=-3$ is the lower boundary of our calculations.}
 \label{fig:Kap0effect}
\end{figure}

\begin{figure}
 \begin{center}
   \includegraphics[width=6cm]{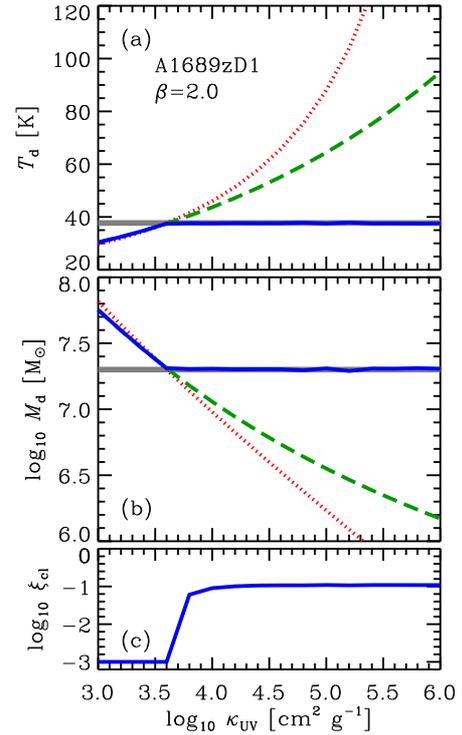}
 \end{center}
 \caption{The same as Figure~\ref{fig:Kap0effect}, but for the UV mass absorption coefficient, $\kappa_{\rm UV}$.}
 \label{fig:KapUVeffect}
\end{figure}

As discussed in section~\ref{subsec:dust}, the mass absorption coefficients are uncertain.
Here we discuss the dependence of the $M_{\rm d}$ and $T_{\rm d}$ estimates on $\kappa_{\rm IR,0}$ and $\kappa_{\rm UV}$.
Figures~\ref{fig:Kap0effect} and \ref{fig:KapUVeffect} show the best-fit values obtained for $M_{\rm d}$ and $T_{\rm d}$ as functions of the mass absorption coefficients $\kappa_{\rm IR,0}$ and $\kappa_{\rm UV}$, respectively.
We also show the clumpiness parameter $\xi_{\rm cl}$ for the clumpy geometry.

For the modified black-body fit, $\kappa_{\rm IR,0}$ does not affect $T_{\rm d}$, because $T_{\rm d}$ is determined only from the shape of the IR SED.
With a fixed $T_{\rm d}$, when $\kappa_{\rm IR,0}$ increases, the IR flux density increases. 
To match the observed IR flux density, $M_{\rm d}$ must therefore decrease. Consequently,
$M_{\rm d}$ is inversely proportional to $\kappa_{\rm IR,0}$, as shown by equation~(\ref{eq:fnu}).
On the other hand, there is no dependence of either $M_{\rm d}$ or $T_{\rm d}$ on $\kappa_{\rm UV}$.
This is shown by the thick grey lines in Figures~\ref{fig:Kap0effect} and \ref{fig:KapUVeffect}.

In the radiative-equilibrium algorithm, the dependencies on $\kappa$ can become non-linear.
With a fixed $\kappa_{\rm UV}$ (i.e. Figure~\ref{fig:Kap0effect}), in the shell and homogeneous geometries (the dotted and dashed lines, respectively), as $\kappa_{\rm IR,0}$ increases, $T_{\rm d}$ decreases, in contrast to the modified black-body case. 
$M_{\rm d}$ also decreases as $\kappa_{\rm IR,0}$ increases, but the dependencies are weaker than the inverse proportionality of the modified black-body case. 
The physical mechanism is explained as follows.
When $\kappa_{\rm IR,0}$ increases, $M_{\rm d}$ must decrease to maintain the IR flux density, as in the modified black-body case.
However, in radiative equilibrium, when $M_{\rm d}$ decreases, the optical depth of the system also decreases.
This makes the total energy absorbed by the dust smaller (eqs.~[\ref{eq:taushe},\ref{eq:tauhom}]), and consequently $T_{\rm d}$ is reduced. 
To maintain the observed flux densities with a lower $T_{\rm d}$, however, requires a larger $M_{\rm d}$.
Therefore, the $M_{\rm d}$ reduction is somewhat alleviated
compared to the modified black-body case.

For the clumpy geometry in radiative equilibrium, the additional parameter controlling clumpiness, $\xi_{\rm cl}$, adjusts $T_{\rm d}$ to the best-fit modified black-body solution, because it provides the best description for the functional shape of the IR SED .
Therefore, $T_{\rm d}$ becomes independent of $\kappa_{\rm IR,0}$, and $M_{\rm d}$ is inversely proportional to $\kappa_{\rm IR,0}$, as in the modified black-body case.
However, at the point where $T_{\rm d}$ becomes equal to that in the homogeneous case, the value of $\xi_{\rm cl}$ disappears (i.e. $\xi_{\rm cl}\to0$).
The clumpy case then deviates from the modified black-body case and follows the homogeneous case.
This behaviour is shown by the solid line in Figure~\ref{fig:Kap0effect}.
Note that the case with $\log_{10}\xi_{\rm cl}=-3$ in panel (c) of Figure~\ref{fig:Kap0effect} is the lower boundary of our calculation; it is essentially the case with $\xi_{\rm cl}\to0$.

With a fixed $\kappa_{\rm IR,0}$ (i.e. Figure~\ref{fig:KapUVeffect}) in the shell and homogeneous geometries, $T_{\rm d}$ increases as $\kappa_{\rm UV}$ increases, because the absorbed radiation energy increases.
To meet the observed IR flux densities in this case, the required $M_{\rm d}$ must therefore decrease.
In the clumpy geometry, the SED shape (or $T_{\rm d}$) can be adjusted to match the best-fit modified black-body function by changing the clumpiness parameter, $\xi_{\rm cl}$.
Therefore, $T_{\rm d}$ and $M_{\rm d}$ become independent of $\kappa_{\rm UV}$, as in the simple modified black-body fit.
This breaks down when $\kappa_{\rm UV}$ becomes too small to retain the best-fit $T_{\rm d}$, and the clumpy case then follows the homogeneous case (i.e. $\xi_{\rm cl}\to0$).

\section{Summary}

To understand the origin of dust in the high-$z$ Universe, it is important to obtain reliable estimates of the dust mass, $M_{\rm d}$.
However, this is often difficult, because the infrared (IR) spectral energy distributions (SEDs) of high-$z$ galaxies are sampled sparsely, and it is common to have only a single detection or upper limit available.
Since it has been believed that the dust temperature $T_{\rm d}$ cannot be estimated in such a situation, a value of $T_{\rm d}$ is routinely assumed, and $M_{\rm d}$ depends on the assumed $T_{\rm d}$.
In this paper, we have proposed a method for estimating both $T_{\rm d}$ and $M_{\rm d}$ simultaneously, even from a single data point in the IR SED, by adopting radiative equilibrium for the dust grains.
By solving the radiative-transfer equation in three simple geometries for the distributions of dust and stars--a thin spherical shell, a homogeneous sphere, and a clumpy sphere--we have obtained analytic formulae for estimating $M_{\rm d}$ and the corresponding $T_{\rm d}$.
We have also applied these new formulae to a normal star-forming dusty galaxy at $z=7.5$, A1689zD1.
For this galaxy, we have reported a new detection of the dust continuum at the rest-frame 90 \micron\ wavelength with ALMA.
In conjunction with the two continuum data points previously reported, the IR SED has thus been sampled at three different wavelengths, making this galaxy an ideal example for testing our algorithm.
Furthermore, we have examined the effects of the UV and IR mass absorption coefficients of the dust on the estimation of $M_{\rm d}$ and $T_{\rm d}$.
Unlike the usual $M_{\rm d}$ estimate with an assumed $T_{\rm d}$, the radiative-equilibrium algorithm couples $M_{\rm d}$ with $T_{\rm d}$, and
the effect of different dust mass absorption coefficients on $M_{\rm d}$ becomes non-linear.

We first performed the usual modified black-body fit to the observed IR SED of A1689zD1 and obtained $T_{\rm d}=38_{-6}^{+9}$~K with the spectral index $\beta=2.0$. 
The corresponding dust mass is $\log_{10}(M_{\rm d}/{\rm M_\odot})=7.3_{-0.3}^{+0.4}$ for a mass absorption coefficient at 100~\micron\ of $\kappa_{\rm IR,0}=30$ cm$^2$ g$^{-1}$.
We have also found that $T_{\rm d}=66$~K and 48~K for $\beta=1$ and 1.5, respectively, which also correspond to $\log_{10}(M_{\rm d}/{\rm M_\odot})=6.5$ and 6.9.
Next, we obtained estimates by adopting radiative equilibrium.
The resulting dust temperatures are $70_{-5}^{+7}$~K and $57_{-2}^{+3}$~K for the thin shell and homogeneous-sphere cases, respectively, assuming $\beta=2.0$.
The corresponding dust masses are $\log_{10}(M_{\rm d}/{\rm M_\odot})=6.5\pm0.1$ and $6.7\pm0.1$, respectively, if $\kappa_{\rm UV}=5.0\times10^4$ cm$^2$ g$^{-1}$ and $\kappa_{\rm IR,0}=30$ cm$^2$ g$^{-1}$.
Even if we use only a single data point from the IR SED, the resulting $T_{\rm d}$ and $M_{\rm d}$ are very similar.
Therefore, our algorithm can be applied to any galaxy that has been observed at least at one IR wavelength.

For the clumpy geometry, we found the same $T_{\rm d}$ and $M_{\rm d}$ as for the modified black-body fitting.
This is because the clumpiness parameter $\xi_{\rm cl}$ adjusts $T_{\rm d}$ to the value that gives the statistically best description of the observed IR data points under the assumed SED shape (i.e. the modified black-body function).
The resulting clumpiness parameter is $\log_{10}\xi_{\rm cl}\simeq-1$.
The definition of the parameter is $\xi_{\rm cl}=\eta_{\rm cl}/f_{\rm cl}$, where $\eta_{\rm cl}$ is the clump-to-system size ratio and $f_{\rm cl}$ is the volume filling factor of the clumps.
If we consider a clump size of $\sim10$ pc, which is the same order as the sizes of giant molecular clouds in our Galaxy \citep{Larson81} and nearby galaxies \citep{Fukui08}, and for a galaxy size of $\sim1$ kpc (see Table~\ref{tab:obsdata}), which is similar to the typical scale of high-$z$ galaxies \citep{Kawamata18}, we find $\eta_{\rm cl}\sim0.01$ and $f_{\rm c}\sim0.1$.
Such $<100$-pc-sized molecular clouds have in fact been observed in a gravitationally lensed high-$z$ galaxy \citep{Tamura15}.
In the future, it will be interesting to compare the inferred clump filling factor to more detailed ISM observations.
If two or more data points are available for the IR SEDs, this kind of discussion about the structure of the ISM becomes possible.
With only a single SED data point, one may assume $\xi_{\rm cl}=0.1$ as a fiducial value.


\section*{Acknowledgements}

The authors thank the anonymous referee for insightful comments which are useful to improve the quality of this paper.
We also thank Hiroyuki Hirashita for discussions about the radiative equilibrium algorithm and Hiroshi Kimura for organizing a series of Cosmic Dust meetings where we were inspired by discussions of cosmic dust in general.
We acknowledge support from NAOJ ALMA Scientific Research Grant number 2016-01 A (A.K.I. and T.H.), JSPS KAKENHI grant 17H01114 (A.K.I.), and Leading Initiative for Excellent Young Researchers, MEXT, Japan (T.H.).
ALMA is a partnership of ESO (representing its member states), NSF (USA) and NINS (Japan), together with NRC (Canada), NSC and ASIAA (Taiwan) and KASI (Republic of Korea), in cooperation with the Republic of Chile. The Joint ALMA Observatory is operated by ESO, AUI/NRAO and NAOJ.
The authors thank Enago (www.enago.jp) for the English language review.

\appendix

\section{Mass absorption coefficients of some dust models}\label{sec:MAC1}

\begin{figure}
    \centering
    \includegraphics[width=6cm]{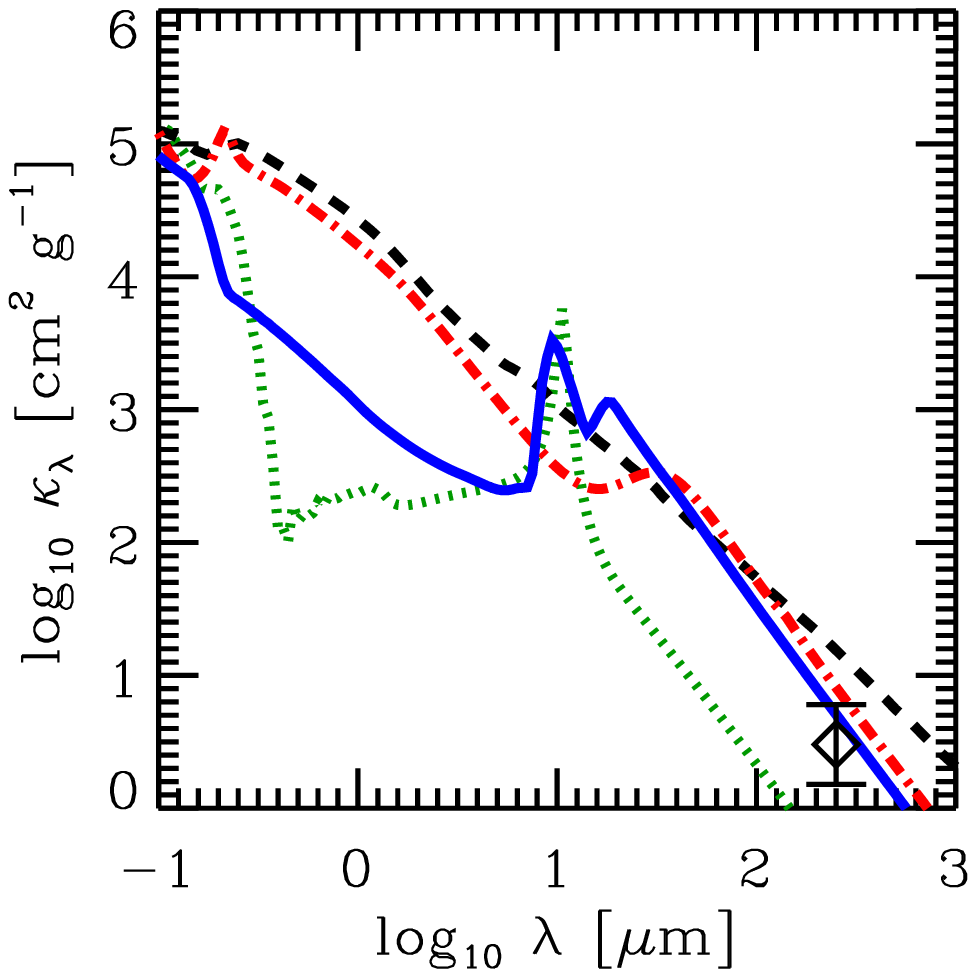}
    \caption{A comparison of mass absorption coefficients for several dust models: smoothed UV astronomical silicates (solid: Draine \& Lee 1984, Laor \& Draine 1993, Weingartner \& Draine 2003), graphite (dot-dashed: Draine \& Lee 1984, Laor \& Draine 1993), amorphous carbon (dashed: Zubko et al.~1996), and SiC (dotted: Laor \& Draine 1993). The so-called MRN size distribution (Mathis et al.~1977) is assumed for all models. The diamond with error-bars is an empirical observational estimate (Bianchi et al.~2019).}
    \label{fig:kappa}
\end{figure}

\begin{figure}
    \centering
    \includegraphics[width=8cm]{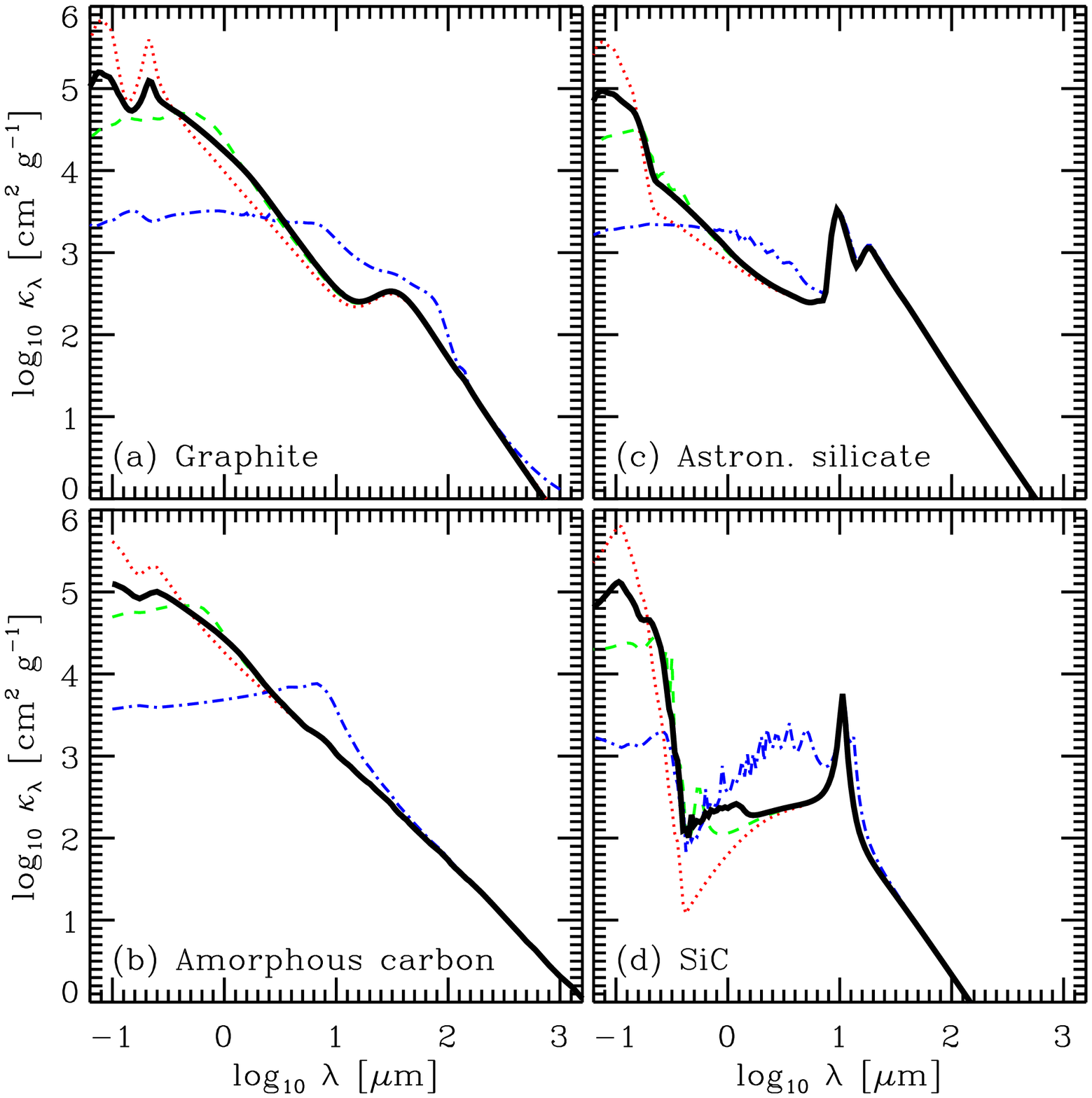}
    \caption{The same as Figure~\ref{fig:kappa}, but showing grain-size effects for four dust models indicated in each panel. The solid lines show averages over the MRN grain-size distribution (Mathis et al.~1977). The dotted, dashed, and dot-dashed lines are single-size cases with dust-grain radii of 0.01~\micron, 0.1~\micron, and 1~\micron.}
    \label{fig:kappa2}
\end{figure}

In Figure~\ref{fig:kappa}, we show the mass absorption coefficients (MACs) of some theoretical dust models integrated over the standard grain size distribution function--the MRN distribution--in the diffuse ISM of the Milky Way \citep{MRN77}.
The assumed maximum and minimum grain radii are 0.25 \micron\ and 0.005 \micron, respectively.
The two carbonaceous models show similar wavelength dependencies.
The silicate and SiC models also share similar trends like the Si-O features at around 10 \micron\ and 18 \micron, as well as transparency in the optical to near-infrared range.
The MAC values at FIR and submillimetre wavelengths are similar to each other, except for the SiC model, and are marginally consistent with the empirical estimate by \cite{Bianchi19}.
Figure~\ref{fig:kappa2} shows the cases of single-size grains of 0.01 \micron, 0.1 \micron, and 1 \micron\ for the same dust models.
It is well known that the MAC values are sensitive to the grain size if the radiation wavelength is comparable to or smaller than the size.
Therefore, the MAC at the UV wavelength shows a large variation depending on the grain size, while that at the FIR wavelength is insensitive to it for the grain sizes considered.
The model dependency on the MAC values at UV wavelengths of 0.1--0.3 \micron\ are modest as listed in Table~\ref{tab:dust} for the cases of the MRN distribution and the single-size of 0.1 \micron.
We have chose $5.0\times10^4$ cm$^2$ g$^{-1}$ as the fiducial value for the UV MAC.\\

\section{Laboratory measurements of the IR emissivity of dust}\label{sec:MAC2}

In this appendix, we present a short review of laboratory measurements of the IR emissivity of cosmic-dust analogues. 
This may be useful in aiding high-$z$ astronomers to understand the current state of this research field.

Dust is classified into carbon types and silicate types and is further classified into amorphous and crystalline.
Especially after the {\it ISO} mission, which obtained beautiful observations of crystalline features (e.g. \citealt{Waters98,Molster02,Honda03}), experimental research on crystalline silicates has been actively conducted.
Table~\ref{tab:labdata} is a summary of laboratory measurements of the mass absorption coefficient (MAC) of some representative silicate materials in the FIR region.
These measurements can be used to determine the properties of circumstellar dust and the physical environments around relatively nearby stars in our Galaxy.
There have also been some reports of detections of crystalline silicates in Ultra-Luminous InfraRed Galaxies (ULIRGs: e.g. \citealt{Spoon06}) beyond our own Galaxy.

It is widely known that the MACs obtained from laboratory work depend strongly on various parameters such as the mineral species, chemical composition, structure, size and shape of particles, temperature, crystallinity, agglomeration, and so on.
As shown in Table~\ref{tab:labdata}, there are many data sets for candidate dust materials from many groups.
However, the measurement conditions are different from each other, and some of the parameters mentioned above are difficult to control in the laboratory. 
Thus, there are few cases where these data are quantitatively consistent.

Many experimental MAC data have been retrieved from measurements of the absorption spectrum of powdered samples.
In the FIR region this method requires a relatively large sample concentration, and it is susceptible to the agglomeration of sample particles.
\citet{Koike94} report that agglomeration makes the MAC larger, even in the FIR region, which
makes the measurements technically difficult.
In addition, many data obtained from powdered samples often present MAC values measured in a dispersing medium (such as KBr for the mid-IR and polyethylene for the FIR measurements).
These MAC values must be converted to the values in vacuum. 
The conversion formula is simple, as $\kappa_{\rm{vac}} = \kappa_{\rm{med}}/{n_{0}}^3$, 
where $\kappa_{\rm{vac}}$, $\kappa_{\rm{med}}$, and $n_{0}$ are the MAC in vacuum, the MAC in the medium, and the refractive index of the medium, respectively (e.g. \citealt{Koike89}).
The medium used for FIR measurement is usually polyethylene, for which the refractive index is about 1.53 at FIR wavelengths \citep{Palik97}.
Moreover, the Bruggeman rule can be applied for a case in which the sample concentration is high, particularly for submillimetre measurements \citep{Mennella98,Boudet05}.
If necessary, the values of the MAC and $\beta$ listed in Table~\ref{tab:labdata} have been corrected for this effect.


On the other hand, some experiments retrieve optical constants (or complex dielectric constants) from reflection measurements from polished, bulk, single crystals 
\citep{Suto06,Sogawa06,Demichelis12,Zeidler15}.
This method, however, is difficult to apply to amorphous materials, because it is difficult to polish them.

Measurements of MACs at low temperatures have also been performed extensively, and the temperature dependence of the MACs has been reported.
It is clear from reports of many groups that the spectral index $\beta$ increases as the temperature decreases, although there is a variance in the absolute value of each group of data 
(e.g. \citealt{Mennella98,Chihara01,Chihara02,Koike03,Boudet05,Koike06,Murata09,Coupeaud11,Mutschke13,Demyk17a,Demyk17b,MutschkeMohr19}).
\citet{Coupeaud11} have investigated the spectral index $\beta$ in the range 100--1000 $\micron$ using interstellar-dust analogues with several chemical compositions.
They report that the MACs of amorphous materials exhibit temperature dependence: a larger $\beta$ for a lower temperature.
Their very important argument regarding $\beta$ is that 
"{\it The spectral shape of the MAC curve does not follow a simple 
asymptotic behaviour with the wavelength so that it cannot be described
correctly with a single spectral index.}"
This has been confirmed again in the experiments by \citet{Demyk17a,Demyk17b}.

In particular, it has been reported that the MAC at low temperatures has a large difference in $\beta$.
\citet{SekiYamamoto80} theoretically expect the wavelength-dependence of the MAC to be proportional to $\lambda^{-2}$ for crystalline and to $\lambda^{-1}$ for amorphous materials.
However, larger values ($\beta=4$--5) have been reported in some data sets \citep{Chihara01,Mutschke13}.
According to \citet{MutschkeMohr19}, there is an interesting report that values of $\beta$ significantly larger than the theoretical prediction may be due to phonon processes in the crystal lattice that behave differently at different temperatures.


\begin{table*}
 \caption{Values of $\kappa$ and $\beta$ for dust analogues from recent laboratory measurements. Amorphous materials are indicated with *.}
 \label{tab:labdata}
 \begin{tabular}{llllll}
 \hline
 & \multicolumn{2}{c}{$\kappa$ (cm$^2$ g$^{-1}$) at 100 \micron} 
    & \multicolumn{2}{c}{Spectral index : $\beta$} & \\
    \cline{2-5}
 Material & Room Temp. & Low Temp. & Room Temp. & Low Temp. & Reference and comments \\
 \hline
Forsterite & 47 (295 K) & 33 (24 K) & 2.04 (295 K) & 2.32 (24 K)    & Mennella et al.~(1998) \\
Fayalite & 39 & 32 & 1.98 & 2.18 & $\kappa$s are converted from the values at 1 mm. \\
Fayalite (amorphous)* & 112 & 94 & 1.35 & 2.04 & $\beta$ measured between 0.1--2 mm. \\ 
\hline
Forsterite & 15 (295 K) & 2 (4 K) & 2.6 (295 K) & 5.0 (4 K) & Chihara et al.~(2001) \\
Orthoenstatite & 25 & 5 & 2.3 & 3.4 & \\ 
\hline
Orthoenstatite & 3.0 & & & & Chihara et al.~(2002) \\
Clinoenstatite & 2.3 & & & & \\ 
\hline
Forsterite & 7.4 (300 K) & 1.48 (4 K) & & & Koike et al.~(2003) \\ 
\hline
Silica monosphare 1.5 \micron * & 20 (300 K) & 12 (10 K) & 0.94 / 1.1 (300 K) & 1.27 / 2.77 (10 K) & Boudet et al.~(2005) \\
Fumed silica* & 13 & 10 & 1.12 / 1.12 & 1.28 / 2.44 & $\kappa$s are taken from their figures~2--5. \\
Enstatite (glass)* & 49 & 41 & 1.52 / 1.58 & 1.58 / 2.14 & $\beta$ measured at 100--200 \micron~ / 0.5--1 mm. \\
Enstatite (sol-gel:amorphous)* & 50 & 40 & 1.68 / 1.44 & 1.92 / 2.74 & \\ 
\hline
HAS (Heated Amorphous Silicate: & 4.44 (300 K) & 0.41 (50 K) & & & Murata et al.~(2009) \\
chondritic enstatite) & & & & & \\ 
\hline
Mg$_{2.05}$SiO$_{4}$ & & & 1.9--2.5 (300 K) & 1.9--4.5(10 K) & Coupeaud et al.~(2011) \\
Mg$_{0.98}$SiO$_{3}$ & & & 0.9--1.7 & 0.9--2.5 & $\beta$ depends on the wavelength and temperature. \\ 
\hline
Fo$_{89}$ (olivine powder) & 13 (300 K) & 6 (10 K) & & & Mutschke et al.~(2013) \\
Fo$_{89}$ (oliine single crystal) & 3.5 & 0.09--0.35 & $\sim$ 2 (300 K) & $>$ 5 (10 K) & \\ 
\hline
Forsterite X35 (glassy silicate)* & 217 (300 K) & 195 (10 K) & $\sim$ 2.2 (300 K) & $\sim$ 3.5 (10 K) & Demyk et al.~(2017a) \\
Enstatiite X50 (glassy silicate)* & 152--160 & 137--153 & 1.6--2 & 2--2.7 (10 K) & MgO : SiO$_2$ = 1-X : X, $\beta$ at 500 \micron \\ 
\hline
E10 (sol-gel:amorphous)* & 305 (300 K) & 260 (10 K) & $\sim$ 2.3 (300 K) & $\sim$ 2.7 & Demyk et al.~(2017b) \\
E20 (sol-gel:amorphous)* & 244 & 225 & $\sim$ 2.1 & $\sim$ 2.9 & MgO : SiO$_2$ = 1-X : X \\
E30 (sol-gel:amorphous)* & 218 & 195 & $\sim$ 1.9 & $\sim$ 2.5 & $\beta$ at 500 \micron \\
E40 (sol-gel:amorphous)* & 245 & 219 & $\sim$ 1.9 & $\sim$ 2.5 &  \\ 
\hline
Forsterite (single crystal) & & 0.009-0.025 & & & Mutschke \& Mohr (2019) \\
Enstatite (single crystal) & 2.4-5 & 0.35-1.3 & & & \\
\hline
 \end{tabular}%
\end{table*}


\bsp	
\label{lastpage}

\begin{thebibliography}{}

\bibitem[Aoyama et al.(2018)]{Aoyama18}
Aoyama, S., Hou, K.-C., Hirashita, H., Nagamine, K., Shimizu, I., 
2018, MNRAS, 478, 4905

\bibitem[Aoyama et al.(2019)]{Aoyama19}
Aoyama, S., Hirashita, H., Lim, C.-F., Chang, Y.-Y., Wang, W.-H., 
Nagamine, K., Hou, K.-C., Shimizu, I., et al., 
2019, MNRAS, 484, 1852

\bibitem[Arata et al.(2019)]{Arata19}
Arata, S., Yajima, H., Nagamine, K., Li, Y., Khochfar, S., 
2019, MNRAS, 488, 2629

\bibitem[Behrens et al.(2018)]{Behrens18}
Behrens, C., Pallottini, A., Ferrara, A., Gallerani, S., Vallini, L., 
2018, MNRAS, 477, 552

\bibitem[Bianchi et al.(2019)]{Bianchi19}
Bianchi, S., Casasola, V., Baes, M., Clark, C. J. R., Corbelli, E., 
Davies, J. I., De Looze, I., De Vis, P., et al., 
2019, A\&A, 631, 102

\bibitem[Boquien et al.(2019)]{Boquien19}
Boquien, M., Burgarella, D., Roehlly, Y., Buat, V., Ciesla, L., 
Corre, D., Inoue, A. K., Salas, H., 2019, A\&A, 622, 103

\bibitem[Boudet et al.(2005)]{Boudet05} 
Boudet, N., Mutschke, H.,Nyral, C., et al., 2005, ApJ, 633, 272

\bibitem[Bradley et al.(2008)]{Bradley08}
Bradley, L. D., Bouwens, R. J., Ford, H. C., Illingworth, G. D., 
Jee, M. J., Ben{\' i}tez, N., Broadhurst, T. J., Franx, M., et al., 
2008, ApJ, 678, 647

\bibitem[Buat et al.(2005)]{Buat05}
Buat, V., Iglesias-P{\'a}ramo, J., Seibert, M., Burgarella, D., 
Charlot, S., Martin, D. C., Xu, C. K., Heckman, T. M., et al.,
2005, ApJ, 619, L51

\bibitem[Calura et al.(2017)]{Calura17}
Calura, F., Pozzi, F., Cresci, G., Santini, P., Gruppioni, C., 
Pozzetti, L., Gilli, R., Matteucchi, F., et al., 
2017, MNRAS, 465, 54

\bibitem[Capak et al.(2015)]{Capak15}
Capak, P. L., Carilli, C., Jones, G., Casey, C. M., Riechers, D., 
Sheth, K., Carollo, C. M., Ilbert, O., et al., 
2015, Nature, 522, 455

\bibitem[Casey(2012)]{Casey12}
Casey, C. M., 2012, MNRAS, 425, 3094

\bibitem[Casey et al.(2018)]{Casey18}
Casey, C. M., Zavala, J. A., Spilker, J., da Cunha, E., Hodge, J.,
Hung, C.-L., Staguhn, J., Finkelstein, S. L., Drew, P., 
2018, ApJ, 862, 77

\bibitem[Chary \& Elbaz(2001)]{CharyElbaz01}
Chary, R., Elbaz, D., 2001, ApJ, 556, 562

\bibitem[Chihara et al.(2001)]{Chihara01} 
Chihara, H., Koike, C., \& Tsuchiyama, A., 2001, PASJ, 53, 243

\bibitem[Chihara et al.(2002)]{Chihara02} 
Chihara, H., Koike, C., Tsuchiyama, A., et al., 2002, A\&A, 391, 267

\bibitem[Clark et al.(2016)]{Clark16}
Clark C. J. R., Schofield S. P., Gomez H. L., Davies J. I., 
2016, MNRAS, 459, 1646

\bibitem[Clark et al.(2019)]{Clark19}
Clark, C. J. R., De Vis, P., Baes, M., Bianchi, S., Casasola, V., 
Cassar{\`a}, L. P., Davies, J. I., Dobbels, W., et al., 
2019, MNRAS, 489, 5256

\bibitem[Coupeaud et al.(2011)]{Coupeaud11} 
Coupeaud, A., Demyk, K., Meny, C., et al., 2011, A\&A, 535, A124

\bibitem[da Cunha et al.(2008)]{daCunha08}
da Cunha, E., Charlot, S., Elbaz, D., 2008, MNRAS, 388, 1595

\bibitem[da Cunha et al.(2013)]{daCunha13}
da Cunha, E., Groves, B., Walter, F., Decarli, R., Weiss, A., 
Bertoldi, F., Carilli, C., Daddi, E., et al., 
2013, ApJ, 766, 13

\bibitem[Dale \& Helou(2002)]{DaleHelou02}
Dale, D. A., Helou, G., 2002, ApJ, 576, 159

\bibitem[Dale et al.(2014)]{Dale14}
Dale, D. A., Helou, G., Magdis, G. E., Armus, L., 
D{\'i}az-Santos, T., Shi, Y., 2014, ApJ, 784, 83

\bibitem[Demichelis et al.(2012)]{Demichelis12} 
Demichelis, R., Suto, H., No\"el, Y., et al., 2012, MNRAS, 420, 147

\bibitem[Demyk et al.(2017a)]{Demyk17a} 
Demyk, K., Meny, C., Lu, X.-H., et al., 2017, A\&A, 600, A123

\bibitem[Demyk et al.(2017b)]{Demyk17b} 
Demyk, K., Meny, C., Leroux, H., et al., 2017, A\&A, 606, A50

\bibitem[Draine \& Lee(1984)]{DL84}
Draine, B. T., Lee, H.-M., 1984, ApJ, 285, 89

\bibitem[Draine \& Li(2007)]{DraineLi07}
Draine, B., Li, A., 2007, ApJ, 657, 810

\bibitem[Dunne et al.(2000)]{Dunne00}
Dunne, L., Eales, S., Edmunds, M., Ivison, R., Alexander, P., 
Clements, D. L., 2000, MNRAS, 315, 115

\bibitem[Faisst et al.(2017)]{Faisst17}
Faisst, A. L., Capak, P. L., Yan, L., Pavesi, R., Riechers, D. A., 
Bari{\v s}i{\'c}, I., Cooke, K. C., Kartaltepe, J. S., et al., 
2017, ApJ, 847, 21

\bibitem[Ferrara et al.(2017)]{Ferrara17}
Ferrara, A., Hirashita, H., Ouchi, M., Fujimoto, S., 
2017, MNRAS, 471, 5018

\bibitem[Fukui et al.(2008)]{Fukui08}
Fukui, Y., Kawamura, A., Minamidani, T., Mizuno, Y., Kanai, Y., Mizuno, N., 
Onishi, T., Yonekura, Y., et al., 2008, ApJS, 178, 56

\bibitem[Graziani et al.(2019)]{Graziani19}
Graziani, L., Schneider, R., Ginolfi, M., Hunt, L. K., Maio, U., 
Glatzle, M., Ciardi, B., 
2019, MNRAS, submitted (arXiv:1909.07388)

\bibitem[Hashimoto et al.(2018)]{Hashimoto18}
Hashimoto, T., Laporte, N., Mawatari, K., Ellis, R. S., Inoue, A. K., 
Zackrisson, E., Roberts-Borsani, G., Zheng, W., et al., 
2018, Nature, 557, 392

\bibitem[Hashimoto et al.(2019)]{Hashimoto19}
Hashimoto, T., Inoue, A. K., Mawatari, K., Tamura, Y., Matsuo, H., 
Furusawa, H., Harikane, Y., Shibuya, T., et al., 
2019, PASJ, 70, in press

\bibitem[Hildebrand(1983)]{Hildebrand83}
Hildebrand, R. H., 1983, QJRAS, 24, 267

\bibitem[Hirashita et al.(2014)]{Hirashita14}
Hirashita, H., Ferrara, A., Dayal, P., Ouchi, M., 
2014, MNRAS, 443, 1704

\bibitem[Hirashita et al.(2017)]{Hirashita17}
Hirashita, H., Burgarella, D., Bouwens, R., 
2017, MNRAS, 472, 4587

\bibitem[Hobson \& Padman(1993)]{Hobson93}
Hobson, M. P., Padman, R., 1993, MNRAS, 264, 161

\bibitem[Honda et al.(2003)]{Honda03}
Honda, M., Kataza, H., Okamoto, Y. K., Miyata, T., 
Yamashita, T., Sako, S., Takubo, S., Onaka, T.,
2003, ApJ, 585, L59

\bibitem[Imara et al.(2018)]{Imara18}
Imara, N., Loeb, A., Johnson, B. D., Conroy, C., Behroozi, P., 
2018, ApJ, 854, 36

\bibitem[Inoue(2005)]{Inoue05}
Inoue, A. K., 2005, MNRAS, 359, 171

\bibitem[Jones et al.(2017)]{Jones17}
Jones, A. P., K{\"o}hler, M., Ysard, N., Bocchio, M., Verstraete, L.,
2017, A\&A, 602, 46

\bibitem[Kawamata et al.(2018)]{Kawamata18}
Kawamata, R., Ishigaki, M., Shimasaku, K., Oguri, M., 
Ouchi, M., Tanigawa, S., 2018, ApJ, 855, 4

\bibitem[Kennicutt(1998)]{Kennicutt98}
Kennicutt, R. C., 1998, ARA\&A, 36, 189

\bibitem[Knudsen et al.(2017)]{Knudsen17}
Knudsen, K. K., Watson, D., Frayer, D., Christensen, L., 
Gallazzi, A., Micha{\l}owski, M. J., Richard, J., Zavala, J., 
2017, MNRAS, 466, 138

\bibitem[Koike et al.(1989)]{Koike89} 
Koike, C., Hasegawa, H., Asada, N., et al., 1989, MNRAS, 239, 127

\bibitem[Koike \& Shibai(1994)]{Koike94} 
Koike, C., \& Shibai, H., 1994, MNRAS, 269, 1011

\bibitem[Koike et al.(2003)]{Koike03} 
Koike, C., Chihara, H., Tsuchiyama, A., et al., 2003, A\&A, 399, 1101

\bibitem[Koike et al.(2006)]{Koike06} 
Koike, C., Mutschke, H., Suto, H., et al., 2006, A\&A, 449, 583

\bibitem[Laporte et al.(2017)]{Laporte17}
Laporte, N., Ellis, R. S., Boone, F., Bauer, F. E., Qu{\'e}nard, D., 
Roberts-Borsani, G. W., Pell{\'o}, R., P{\'e}rez-Fournon, I., et al., 
2017, ApJ, 837, L2

\bibitem[Larson(1981)]{Larson81}
Larson, R. B., 1981, MNRAS, 194, 809

\bibitem[Liang et al.(2019)]{Liang19}
Liang, L., Feldmann, R., Kere{\v s}, D., Scoville, N. Z., 
Hayward, C. C., Faucher-Gigu{\`e}re, C.-A., Schreiber, C., Ma, X., et al., 
2019, MNRAS, 489, 1397

\bibitem[Ma et al.(2019)]{Ma19}
Ma, X., Hayward, C. C., Casey, C. M., Hopkins, P. F., Quataert, E., 
Liang, L., Faucher-Gigu{\`e}re, C.-A., Feldmann, R., et al., 
2019, MNRAS, 487, 1844

\bibitem[Marrone et al.(2018)]{Marrone18}
Marrone, D. P., Spilker, J. S., Hayward, C. C., Vieira, J. D., 
Aravena, M., Ashby, M. L. N., Bayliss, M. B., B{\'e}thermin, M., et al., 
2018, Nature, 553, 51

\bibitem[Mathis et al.(1977)]{MRN77}
Mathis, J. S., Rumpl, W., Nordsieck, K. H.,
1977, ApJ, 217, 425

\bibitem[McMullin et al.(2007)]{McMullin07}
McMullin, J. P., Waters, B., Schiebel, D., Young, W., Golap, K., 
2007, ASPC, 376, 127

\bibitem[Mennella et al.(1998)]{Mennella98} 
Mennella, V., Brucato, J. R., Colangeli, L., et al., 1998, ApJ, 496, 1058

\bibitem[Micha{\l}owski(2015)]{Michalowski15}
Micha{\l}owski, M. J., 2015, A\&A, 577, 80

\bibitem[Molster et al.(2002)]{Molster02}
Molster, F. J., Waters, L. B. F. M., Tielens, A. G. G. M.,
2002, A\&A, 382, 222

\bibitem[Murata et al.(2009)]{Murata09}
Murata, K., Chihara, H., Koike, C., et al., 2009, ApJ, 698, 1903

\bibitem[Meurer et al.(1999)]{Meurer99}
Meurer, G. R., Heckman, T. M., Calzetti, D., 1999, ApJ, 521, 64 

\bibitem[Mutschke et al.(2013)]{Mutschke13} 
Mutschke, H., Zeidler, S. \& Chihara, H., 2013, EPS, 65, 1139

\bibitem[Mutschke \& Mohr(2019)]{MutschkeMohr19} 
Mutschke, H. \& Mohr, P., 2019, A\&A, 625, A61

\bibitem[Narayanan et al.(2018)]{Narayanan18}
Narayanan, D., Dav{\'e}, R., Jonson, B. D., Conroy, C., Geach, J., 
2018, MNRAS, 474, 1718

\bibitem[Neufeld(1991)]{Neufeld91}
Neufeld, D. A., 1991, ApJ, 370, L85

\bibitem[Nozawa et al.(2003)]{Nozawa03}
Nozawa, T., Kozasa, T., Umeda, H., Maeda, K., Nomoto, K.,
2003, ApJ, 598, 785

\bibitem[Oke \& Gunn(1983)]{OkeGunn83}
Oke, J. B., Gunn, J., 1983, ApJ, 266, 713

\bibitem[Okumura et al.(1996)]{Okumura96}
Okumura, K., Hiromoto, N., Okuda, H., Shibai, H., Nakagawa, T., 
Makiuti, S., Matsuhara, H., 1996, PASJ, 48, L123

\bibitem[Osterbrock(1989)]{Osterbrock89}
Osterbrock, D. E., 1989, Astrophysics of gaseous nebulae and active galactic nuclei (Mill Valley: University Science Books) Appendix 2

\bibitem[Ouchi et al.(2009)]{Ouchi09}
Ouchi, M., Ono, Y., Egami, E., Saito, T., Oguri, M., McCarthy, P. J., 
Farrah, D., Kashikawa, N., et al., 2009, ApJ, 696, 1164

\bibitem[Palik(1997)]{Palik97}
Palik, E.D., 1997, Handbook of Optical Constants of Solids, Academic Press Books, 
Elsevier, Cambridge

\bibitem[Riechers et al.(2013)]{Riechers13}
Riechers, D. A., Bradford, C. M., Clements, D. L., Dowell, C. D., 
P{\'e}rez-Fournon, I., Ivison, R. J., Bridge, C., Conley, A., et al., 
2013, Nature, 496, 329

\bibitem[Rieke et al.(2009)]{Rieke09}
Rieke, G. H., Alonso-Herrero, A., Weiner, B. J., 
P{\'e}rez-Gonz{\'a}lez, P. G., Blaylock, M., Donley, J. L., Marcillac, D., 
2009, ApJ, 692, 556

\bibitem[Seki \& Yamamoto(1980)]{SekiYamamoto80} 
Seki, J. \& Yamamoto, T., 1980, Ap\&SS, 72, 79

\bibitem[Sogawa et al.(2006)]{Sogawa06} 
Sogawa, H., Koike, C., Chihara, H., et al., 2006, A\&A, 451, 357

\bibitem[Spoon et al.(2006)]{Spoon06} 
Spoon, H. W. W., Tielens, A. G. G. M., Armus, L., et al., 2006, ApJ, 638, 759

\bibitem[Struble \& Rood(1999)]{Struble99}
Struble, M. F., Rood, H. J., 1999, ApJS, 125, 35

\bibitem[Suto et al.(2006)]{Suto06} 
Suto, H., Sogawa, H., Tachibana, S., et al., 2006, MNRAS, 370, 1599

\bibitem[Takeuchi et al.(2012)]{Takeuchi12}
Takeuchi, T. T., Yuan, F.-T., Ikeyama, A., Murata, K. L., Inoue, A. K.,
2012, ApJ, 755, 144

\bibitem[Tamura et al.(2015)]{Tamura15}
Tamura, Y., Oguri, M., Iono, D., Hatsukade, B., Matsuda, Y., Hayashi, M.,
2015, PASJ, 67, 72

\bibitem[Tamura et al.(2018)]{Tamura18}
Tamura, Y., Mawatari, K., Hashimoto, T., Inoue, A. K., Zackrisson, E., 
Christensen, L., Binggeli, C., Matsuda, Y., et al., 
2018, ApJ, 874, 27

\bibitem[Totani \& Takeuchi(2002)]{TotaniTakeuchi02}
Totani, T., Takeuchi, T. T., 2002, ApJ, 570, 470

\bibitem[V{\'a}rosi \& Dwek(1999)]{Varosi99}
V{\'a}rosi, F., Dwek, E., 1999, ApJ, 523, 265

\bibitem[Waters et al.(1998)]{Waters98}
Waters, L. B. F. M., Beintema, D. A., Zijlstra, A. A., de Koter, A., 
Molster, F. J., Bouwman, J., de Jong, T., Pottasch, S. R., et al.,
1998, A\&A, 331, L61

\bibitem[Watson et al.(2015)]{Watson15}
Watson, D., Christensen, L., Knudsen, K. K., Richard, J., 
Gallazzi, A., Micha{\l}owski, M., J., 2015, Nature, 519, 327

\bibitem[Weingartner \& Draine(2001)]{WD01}
Weingartner, J. C., Draine, B. T., 2001, ApJ, 548, 296

\bibitem[Zeidler et al.(2015)]{Zeidler15} 
Zeidler, S., Mutschke, H. \& Posch, T., 2015, ApJ, 798, 125

\bibitem[Zubko et al.(1996)]{Zubko96}
Zubko, V. G., Mennella, V., Colangeli, L., Bussoletti, E., 
1996, MNRAS, 282, 1321

\end{thebibliography}
\end{document}